\documentclass{article}

\usepackage{graphicx}%
\usepackage{multirow}%
\usepackage{amsmath,amssymb,amsfonts}%
\usepackage{amsthm}%
\usepackage{mathrsfs}%
\usepackage[title]{appendix}%
\usepackage{xcolor}%
\usepackage{textcomp}%
\usepackage{manyfoot}%
\usepackage{booktabs}%
\usepackage{algorithm}%
\usepackage{algorithmicx}%
\usepackage{algpseudocode}%
\usepackage{listings}%
\usepackage[colorlinks=true,linkcolor=blue,citecolor=blue,urlcolor=blue]{hyperref}

\usepackage[font=small,labelfont=bf]{caption}

\theoremstyle{thmstyleone}%
\theoremstyle{thmstyletwo}%

\theoremstyle{thmstylethree}%

\begin{document}

\title{On Hydrodynamic Formulations of Quantum Mechanics and the Problem of Sparse Ontology}

\author{
Aric Hackebill\thanks{aric.hackebill@uvm.edu}
\and
Bill Poirier\thanks{bill.poirier@uvm.edu}
}

\date{
Department of Chemistry, University of Vermont, Burlington, VT 05405, USA
}

\maketitle

\begin{abstract}
Hydrodynamic reformulations of the Schrödinger equation suggest an interpretation of quantum mechanics in terms of a fluid flowing on configuration space. In the discrete hydrodynamic view, this fluid is not fundamental but emerges from many underlying microscopic fluid components whose collective behavior reproduces quantum phenomena. The most developed realization of this idea is the discrete many interacting worlds (MIW) framework, in which discrete particle-like worlds interact via inter-world forces and quantum probabilities are grounded in direct world counting. But there is also an older, continuous version of MIW. After reviewing the hydrodynamic and MIW formalisms, and emphasizing some of their interpretational advantages over the Everettian Many Worlds and Bohmian approaches, we argue that all discrete hydrodynamic models face a generic structural difficulty, which we call the \emph{problem of sparse ontology}. Because wavefunctions typically branch under decoherence, the discrete components of the fluid are repeatedly partitioned into sub-ensembles, thereby thinning their density in configuration space and driving the dynamics away from the quantum regime once the components become sufficiently sparse. We conclude that successful hydrodynamic completions of quantum mechanics plausibly require an essentially continuous ontology.
\end{abstract}

\section{Introduction}\label{sec1}

In 1926, Madelung introduced an alternative to the standard formulation of quantum mechanics for single-particle systems by recasting the Schr\"odinger equation into a set of equations resembling the continuity and Euler equations of classical fluid dynamics \cite{madelungQuantentheorieHydrodynamischerForm1927,bohmSuggestedInterpretationQuantum1952,Bohm2}. This view has since undergone many stages of development, and there are now variants of Madelung's initial theory that can account for a wide range of quantum phenomena.\footnote{The multi-particle extension is immediate, and can be found in most introductory texts \cite{hollandQuantumTheoryMotion1993,wyattQuantumDynamicsTrajectories2005}; a nonrelativistic spin extension was developed in  \cite{takabayasiHydrodynamicalRepresentationNonRelativistic1954}; relativistic extensions based on the Dirac theory were developed in \cite{takabayasiRelativisticHydrodynamicsDirac1957,fabbriDiracTheoryHydrodynamic2023}.} Central to each of these variants is the formulation of their fundamental dynamical laws in terms of hydrodynamic variables that ostensibly describe the motion of a fluid on a high-dimensional configuration space $\mathcal{Q}$. This immediately raises a question about the nature of the Madelung fluid. 

In a straightforward interpretation of the dynamical equations, the fluid is a physical substance with a genuine (configuration space) matter density and velocity field describing its matter flow. Pushing the analogy with classical fluids further, the fluid can be seen as being composed of many discrete microscopic components (e.g., particles) whose interactions yield fluid-like properties at macroscopic scales. On this view, which we call the \emph{discrete hydrodynamic view} (DHV), the hydrodynamic equations are effective rather than fundamental, and quantum phenomena emerge when large numbers of microscopic components, interacting under a more fundamental microphysical law, are tightly aggregated. 

Alternatively, one may instead take the Madelung fluid to be fundamentally continuous. In this \emph{continuous hydrodynamic view} (CHV), the hydrodynamic fields are taken to represent a continuous fluid evolving on $\mathcal{Q}$, without commitment to an underlying discrete ontology.

Taken on their own, however, both the DHV and CHV amount to bare ontological commitments that do not, by themselves, furnish the hydrodynamic view with enough structure to count as a complete foundational theory. In particular, a bare hydrodynamic ontology does not by itself explain how definite measurement outcomes arise from an extended fluid on $\mathcal{Q}$, how probabilities should be grounded, or how the empirical content of quantum systems can be recovered from the underlying dynamics.

A particularly well-developed family of hydrodynamic theories is provided by the many interacting worlds (MIW) framework, in which the Madelung fluid is modeled as an ensemble of interacting configuration-space point trajectories. It is common for each trajectory to be called a \emph{world}, since the instantaneous position of a trajectory in $3N$-dimensional $\mathcal{Q}$ encodes the behavior of $N$ particles moving in a three-dimensional space. In its original and continuous formulation \cite{poirierBohmianMechanicsPilot2010,poirier11nowaveCCP6,schiffCommunicationQuantumMechanics2012,PARLANT20123,rqtpap,poirier14prx,LiCaH,poirier20relsym,poirier21fermi,poirier23nuclearoptpot,lombardiniInteractingQuantumTrajectories2024,poirier24spin2}, which we refer to as the \emph{continuous many interacting worlds} (CMIW) approach, the fluid is composed of a continuum of such worlds, with one world passing through each point of $\mathcal{Q}$ at any instant in time. In this picture, the hydrodynamic fields arise as smooth collections of the continuum of worlds, making the CMIW framework a particular instantiation of the CHV.\footnote{A closely related ``continuum-of-worlds'' formulation has also been developed by Bostr{\"o}m~\cite{bostrom2015}, in which $|\psi|^2$ is interpreted as a density of worlds distributed over $\mathcal{Q}$ and Born probabilities are derived by applying the Laplacian rule of indifference to the resulting world-volume measure.} 

There is also a discrete variant of MIW, \emph{discrete many interacting worlds} (DMIW), in which the Madelung fluid is composed of a finite or countably infinite collection of interacting worlds \cite{hallQuantumPhenomenaModeled2014,herrmannEigenstatesManyInteracting2023}. In this case, the hydrodynamic fields function as smooth, coarse-grained approximations to the fine-grained microbehavior of a dense ensemble of worlds. In the high-world-density regime, these discrete world-trajectories closely approximate the Bohmian trajectories, while at a fundamental level, their dynamics are governed by explicit inter-world interaction laws. The DMIW theory naturally fits in as a concrete realization of the DHV.

Importantly, MIW frameworks supply much of the interpretational structure that the bare DHV and CHV leave unresolved. In Sect.~\ref{sec3}, we illustrate this for the discrete case, and show how the methods used in DMIW can be generalized to the wider DHV. Along the way, we point out several interpretational advantages that the DMIW theory, and by extension the DHV, enjoys over competing approaches in quantum foundations. Among these is the relative ease with which DMIW accounts for quantum probabilities by grounding them in direct world counting. This obviates the need for complex decision-theoretic postulates that some Everettians appeal to when direct branch counting fails to yield the correct Born weights.\footnote{For an extended defense of the decision-theoretic approach, see Wallace’s discussion of probability in the Everett interpretation \cite[Ch.~8]{wallaceEmergentMultiverseQuantum2012}.} It also partially avoids the need to introduce typicality \cite{durrQuantumEquilibriumOrigin1992} or dynamical relaxation \cite{valentiniSignallocalityHiddenvariablesTheories2002} arguments to explain the mysterious dual nature of the wavefunction in Bohmian mechanics, where it both guides particles along their trajectories and simultaneously encodes the probability distribution of particles in the universe---two features that, conceptually at least, need not be so tightly connected.

Despite these advantages, the primary aim of this work is to point out a potential difficulty with the DHV that stems from its inherently discrete nature. The issue can be appreciated at a glance by noticing the tendency of quantum wavefunctions to split into decoherent branches in the presence of a broad range of interactions, which, from the perspective of the DHV, corresponds to the Madelung fluid dividing into distinct regions of $\mathcal{Q}$. Such splitting events necessarily cause the components of the fluid to separate from each other in $\mathcal{Q}$, thereby thinning out the fluid's density. After many such thinning events, the components may become so sparse and distant from one another that they cease to interact with each other significantly---leading to a breakdown in the hydrodynamic description and, consequently, to unacceptably large deviations from standard quantum predictions.  

In the main body of this paper, we consider several cases where departures from standard quantum mechanics might arise. In Sect. \ref{sec4} we introduce a DMIW model for a single spin-$\frac{1}{2}$ particle, and in Sect. \ref{sparseworldargument} we show how its predictions deviate from standard quantum mechanics when the Madelung fluid has been thinned out by a sequence of concatenated Stern-Gerlach analyzers. In Sect. \ref{general} we generalize the argument and show that any DHV theory that (i) grounds probabilities in direct world counting and (ii) has configuration-space-local (\(\mathcal{Q}\)-local) fundamental interactions (formal definition in Sect.~\ref{DMIWdynamics}) will exhibit such unacceptably large deviations from standard quantum mechanics, due to the induced sparsity of its components. Since both features seem essential to the DHV, we conclude that the DHV is generally afflicted by this \emph{problem of sparse ontology}. 

\section{The Hydrodynamic View}\label{sec2}

The hydrodynamic equations for a non-relativistic system of $N$ spinless particles, characterized by the configuration space location $\textbf{x}=(\vec{x}_1,...,\vec{x}_N)$, evolving under a scalar potential $V(\textbf{x},t)$, can be derived by first substituting the polar form of the wavefunction, 

    \begin{equation}\label{polar}  
        \psi(\textbf{x},t)=R(\textbf{x},t)e^{iS(\textbf{x},t)},
    \end{equation}

\noindent (with $R$ and $S$ both real-valued) into the many-body Schr\"odinger equation \cite{madelungQuantentheorieHydrodynamischerForm1927,		bohmSuggestedInterpretationQuantum1952,Bohm2,takabayasiHydrodynamicalRepresentationNonRelativistic1954,durrBohmianMechanicsMeaning1995,sebensQuantumMechanicsClassical2015,hollandQuantumTheoryMotion1993,wyattQuantumDynamicsTrajectories2005}:

    \begin{equation}
        i\hbar\frac{\partial{\psi}(\textbf{x},t)}{\partial{t}}=\displaystyle\sum_{i=1}^N\frac{-\hbar^2}{2m_i}\nabla_i^2\psi(\textbf{x},t)+V(\textbf{x},t)\psi(\textbf{x},t)
    \end{equation}

\noindent The result is then recast completely in terms of hydrodynamic variables, by replacing $R(\textbf{x},t)$ and $S(\textbf{x},t)$ with the fluid density, $\rho(\textbf{x},t)$, and the Bohmian velocity field, $\textbf{v}(\textbf{x},t)$, which are given by 

    \begin{align}
        \rho(\textbf{x},t)&=\lvert\psi(\textbf{x},t)\rvert^2=R^2(\textbf{x},t)\label{density},
        \\
        \textbf{v}(\textbf{x},t)&=(\vec{v}_1(\textbf{x},t),\ldots,\vec{v}_N(\textbf{x},t))
        \\
        &=\left(\frac{\hbar}{m_1}\vec{\nabla}_1S(\textbf{x},t),\ldots,\frac{\hbar}{m_N}\vec{\nabla}_NS(\textbf{x},t)\right).\label{velocityfield}
    \end{align}

\noindent After separately equating real and imaginary parts, and re-expressing in terms of $\rho(\textbf{x},t)$ and $\textbf{v}(\textbf{x},t)$, the hydrodynamic equations
 in component form are given by\footnote{A detailed derivation of these \emph{precise} equations is given in \cite[Sect.~4]{sebensQuantumMechanicsClassical2015}.}

    \begin{align}
        m_j\frac{d\vec{v}_j(\textbf{x},t)}{dt}
          &= -\vec{\nabla}_j\left\{\sum_{i=1}^N \frac{-\hbar^2}{2m_i}\left[
             \frac{\nabla_i^2 \sqrt{\rho(\textbf{x},t)}}{\sqrt{\rho(\textbf{x},t)}}\right]
             + V(\textbf{x},t)\right\}\label{Euler3}, \\
             \frac{\partial\rho(\textbf{x},t)}{\partial{t}}
         &= -\sum_{i=1}^N
            \vec{\nabla}_i \cdot (\rho(\textbf{x},t) \vec{v}_i(\textbf{x},t) )\label{continuity3}.
    \end{align}  

 The derivation of the hydrodynamic equations outlined here very nearly follows Bohm's original derivation of his trajectory equation \cite{bohmSuggestedInterpretationQuantum1952}. However, the hydrodynamic view differs significantly from the Bohmian view in how it interprets Eqs. (\ref{Euler3}) and (\ref{continuity3}). In the Bohmian view, Eq. (\ref{Euler3}) is meant to describe a single trajectory, $\textbf{x}_B(t)=(\vec{x}_{B_1}(t),\ldots,\vec{x}_{B_N}(t))$, in $\mathcal{Q}$, which is realized when the field expressions in Eq. (\ref{Euler3}) are evaluated at the actual configuration of particles in our universe. As was briefly discussed in Sect. \ref{sec1}, we will refer to a single point trajectory in $\mathcal{Q}$ as a \emph{world} since its evolution in $3N$-dimensional configuration space encodes the motion of $N$ particles in a three-dimensional space.\footnote{One may either (i) regard the \(3N\)-dimensional configuration space as
fundamental, with three-dimensional space emergent \cite{albertElementaryQuantumMetaphysics1996}, or (ii) take three-dimensional space as fundamental and treat $\mathcal{Q}$ as calculational \cite{alloriCommonStructureBohmian2008}. The problem of sparse ontology does not ultimately depend on this choice. In case (i), ``sparsening'' is just the separation of fluid components
in $\mathcal{Q}$. In case (ii), the dynamics are either explicitly written in configuration-space variables and then projected into the three-dimensional space, or they are isomorphic to a configuration-space theory. In both cases, once the component density is low, effective inter-component interactions become negligible and the dynamics deviate radically from standard quantum predictions. The sparse ontology problem is therefore dynamical, not merely
representational. We will refrain from adopting a stance as to which space is fundamental, but we will conceptualize our arguments in the configuration space picture under the assumption that we can translate to the 3D picture at any time without undermining our argument.} Eq. (\ref{Euler3}) then serves as the quantum mechanical extension of Newton's second law, with deviations from classical mechanics explicitly arising due to the presence of the quantum potential,

     \begin{equation}\label{Qpotential}
        Q(\textbf{x},t)=-\sum_{i=1}^N \frac{\hbar^2}{2m_i}\left[
             \frac{\nabla_i^2 \sqrt{\rho(\textbf{x},t)}}{\sqrt{\rho(\textbf{x},t)}}\right].
    \end{equation}

\noindent The field $\rho(\textbf{x},t)$ plays a dual role. It primarily contributes to the guidance of the Bohmian world by entering the quantum potential in Eq. (\ref{Euler3}). However, it is also taken to be a probability density, expressing the likelihood that the Bohmian world is located in a particular region of $\mathcal{Q}$. 

In the hydrodynamic view, the field variables $\rho(\textbf{x},t)$ and $\textbf{v}(\textbf{x},t)$ are taken to describe the properties of an extended fluid, with Eqs. (\ref{Euler3}) and (\ref{continuity3}) serving as the quantum mechanical analogs of the Euler and continuity equations of classical fluid mechanics. The DHV pushes the analogy with classical fluid dynamics further by positing many discrete microscopic components that, when aggregated together, reproduce the behavior of a fluid. In this interpretation, $\textbf{v}(\textbf{x},t)$ describes the average motion of components local to the configuration point $\textbf{x}$, while $\rho(\textbf{x},t)$ is a smooth approximation of the density of components contained in a small region centered around $\textbf{x}$. 

 The DHV, understood in this general sense, allows for a broad class of underlying fluid components. They could, in principle, be point-like objects following temporally continuous (non-jumping) trajectories (i.e., worlds), discrete flashes that do not persist in time,\footnote{Here we have in mind the flash ontology proposed by J. Bell in \cite[Ch.~22]{bellSpeakableUnspeakableQuantum2004}} extended objects such as fields or localized blobs, or even more complex composite structures. In other words, the DHV is compatible with any discrete ontology whose aggregate behavior reproduces the effective hydrodynamic fields in the high-density regime. The DMIW framework represents a concrete realization of this general idea, in which the fluid components are taken to be discrete worlds evolving along temporally continuous paths. What distinguishes the DMIW from the more general DHV is that the former's concrete ontology and explicit dynamical laws are sufficiently well defined to address the central foundational problems of quantum mechanics.

   \begin{figure}
        \centering
        \includegraphics[width=0.7\linewidth]{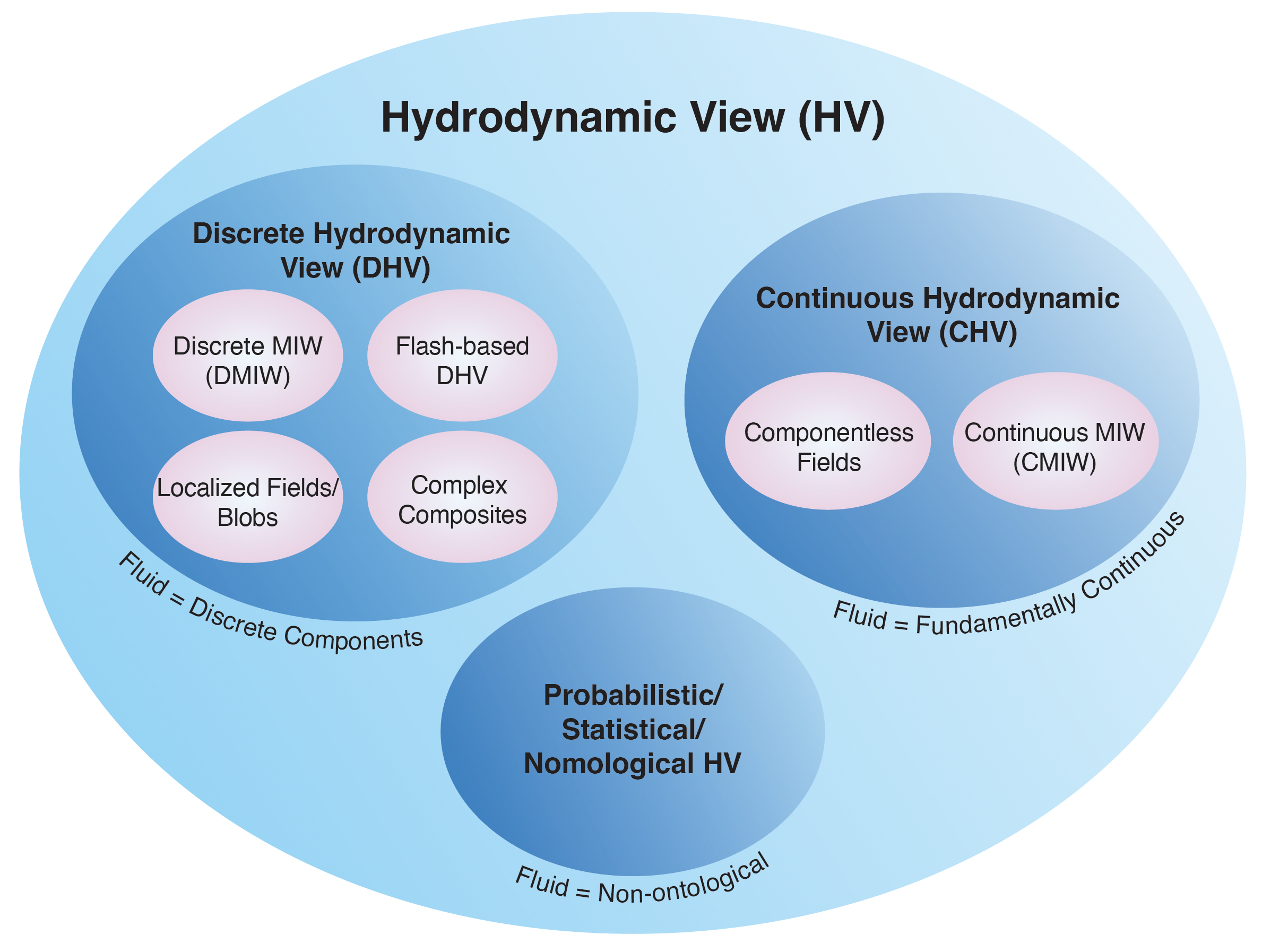}
        \caption{Taxonomy of hydrodynamic interpretations with representative examples.}
        \label{heirarchy}
    \end{figure}
 
 In contrast to the DHV, the CHV takes the Madelung fluid to be fundamentally continuous in nature. In this view, the hydrodynamic fields may be taken as primitive elements of the ontology, rather than as mereological composites of smaller components. Alternatively, they may be conceived as arising from an underlying continuum of fluid components whose motion is exactly captured by the hydrodynamic equations.\footnote{One advantage of continuous hydrodynamic formulations is that their dynamical laws are exact rather than effective.} A concrete realization of this latter perspective is provided by the CMIW framework, in which the Madelung fluid is modeled as a continuum of interacting worlds, each subject to a well-defined and localized inter-world force. As in the discrete case, CMIW fills in the interpretational details that the CHV, taken on its own as a bare ontological claim, leaves unspecified.  

 Finally, one might also imagine hydrodynamic views that do not seek to ascribe ontological significance to the hydrodynamic fields at all. These might include purely probabilistic or statistical interpretations in which the hydrodynamic fields function solely as instrumental tools for computing measurement statistics, or they may be nomological views, in which the hydrodynamic fields play a law-like role governing the motion of some separate supplementary ontology.\footnote{Here we have in mind analogous Bohmian views in which the wavefunction is treated as a law-like ingredient governing the motion of the Bohmian particles; see, e.g., \cite{neyWaveFunctionEssays2013, durrBohmianMechanicsMeaning1995}.} We mention these only to complete the taxonomy of hydrodynamic views. Since our focus is exclusively on hydrodynamic views that posit the fluid, or its constituents, as physically real, we will set this class of interpretations aside.
 
 Having established a taxonomy of hydrodynamic views (see Fig. \ref{heirarchy}), we now direct our attention to the DHV, which is the central target of our argument. As it stands, the DHV is only a bare ontological claim and does not yet possess the interpretational resources needed to function as a complete foundational theory. To put the DHV on equal footing with the other major programs in quantum foundations, we must show how it accounts for quantum phenomena while addressing standard foundational problems.\footnote{For an excellent discussion of how these challenges can be met, see Sebens’s account of \emph{Newtonian QM}, in which the fluid components are taken to be worlds \cite{sebensQuantumMechanicsClassical2015}.} In the next section, we therefore examine the DMIW framework as a fully worked-out realization of the DHV, and show how its techniques can be generalized to the broader class of DHV models. Once the DHV is completed in this way, we will be in a position to assess its viability as a foundational theoretical framework. 

\section{DMIW in Detail}\label{sec3}

 According to DMIW theory, the Madelung fluid is composed of many interacting worlds, which move along continuous trajectories and are discretely distributed in $\mathcal{Q}$. 
Let the Madelung fluid describing a system of $N$ particles be composed of $K$ worlds\footnote{More precisely, we assume local finiteness: every bounded region of $\mathcal{Q}$ contains only finitely many worlds. Hence, the total ensemble may be countably infinite. That said, the creators of discrete DMIW themselves have expressed a preference for an ontology where the \emph{total} number of worlds is finite (personal communication).} whose trajectories are written as 

    \begin{equation}\label{world}
        \mathbf{x}_k(t) =
            \begin{pmatrix}
                \vec{x}_{k_1}(t) \\[2pt]
                \vec{x}_{k_2}(t) \\[2pt]
                \vdots \\[2pt]
                \vec{x}_{k_N}(t)
            \end{pmatrix}
        \in \mathbb{R}^{3N},
    \end{equation}

\noindent where  $k\in\{1,\ldots,K\}$ is an index over worlds and $n\in\{1,\ldots,N\}$ is an index over particles. The dynamics of each world enter as a modification to Newton's Second Law 

    \begin{equation}\label{WisemanTrajectory}
        \mathbb{M}\,\ddot{\mathbf{x}}_{k}=-\nabla_{\textbf{x}_k}(V(\textbf{x}_k)+U(\textbf{x}_1,\ldots,\textbf{x}_K)),
    \end{equation}

\noindent where $\mathbb{M}=\mathrm{diag}(m_1 I_3,\ldots,m_N I_3)$ encodes the masses of the $N$ particles, $V$ is the classical potential, and $U$ is an additional potential that encodes inter-world interactions. Thus, in the DMIW framework, deviations from the classical theory arise purely from the presence of these inter-world interactions.

\subsection{Dynamics and Locality}\label{DMIWdynamics}

The exact dynamics of a DMIW system will depend on the specific form of the inter-world interaction potential that is presumed. In \cite{hallQuantumPhenomenaModeled2014}, Hall et al. introduced a toy model for a single particle in one dimension where the Madelung fluid, which flows in a one-dimensional configuration space, is constituted by an ensemble of worlds whose coordinate labels, $\textbf{x}_k=x_k$, are chosen such that $x_1<x_2<...<x_K$ 
(where this ordering is presumed to be preserved over time). The worlds are taken to interact only with their nearest neighbors, according to\footnote{Note that the ``edge trajectories'' require special consideration in this scheme.}

    \begin{equation}\label{WisemanPotential}
        U(x_1,...,x_K)=\frac{\hbar^2}{8m}\displaystyle\sum_{k=1}^K\left[\frac{1}{x_{k+1}-x_k}-\frac{1}{x_k-x_{k-1}}\right]^2,
    \end{equation}

\noindent which when plugged into Eq. (\ref{WisemanTrajectory}), yields

    \begin{equation}\label{DMIWNSL}
        m\ddot{x}_k=-\frac{\partial{V(x_k)}}{\partial{x_k}}+F_{Q_k}.
    \end{equation}

\noindent Here, $F_{Q_k}$ is an inter-world force acting on the $k$th world, given by

    \begin{equation}\label{DiscreteTrajectory}
        \begin{aligned}
            F_{Q_k}=&\frac{\hbar^2}{4m}\Bigg\{\frac{1}{(x_{k+1}-x_k)^2}\left[\frac{1}{x_{k+2}-x_{k+1}}-\frac{2}{x_{k+1}-x_{k}}+\frac{1}{x_{k}-x_{k-1}}\right]
            \\
            &-\frac{1}{(x_{k}-x_{k-1})^2}\left[\frac{1}{x_{k+1}-x_{k}}-\frac{2}{x_{k}-x_{k-1}}+\frac{1}{x_{k-1}-x_{k-2}}\right]\Bigg\}.
        \end{aligned}
    \end{equation}

$F_{Q_k}$exhibits some interesting properties. It is repulsive in nature, which prevents the world trajectories from crossing, thus preserving the order of the coordinate labels of each world as it evolves in time. It loosely resembles Newton's gravitational law in the sense that the inter-world forces are instantaneous, action-at-a-distance-type influences that are unmediated by local fields. But here, crucially, the ``action'' that we are describing is across $\mathcal{Q}$.\footnote{It is worth noting that the DMIW framework is unusual among other foundational approaches in that its fundamental dynamics are nonlocal in $\mathcal{Q}$. Even the most ``non-local'' (in the usual position-space sense) potential $V(\textbf{x}_k)$ that one can possibly conjure for use in Eq.~(\ref{WisemanTrajectory}), still satisfies a stringent form of locality on $\mathcal{Q}$, in that it applies \emph{only} to the single world $\textbf{x}_k$.} The inter-world interactions do have a (somewhat) local character in the sense that they decay (fairly) rapidly when the separation between the worlds is large. 

For a system of many, densely packed worlds, the trajectories given by Eq. (\ref{DiscreteTrajectory}) closely approximate the Bohmian paths; however, they begin to deviate from the Bohmian paths when the worlds are sparsely distributed and are separated far enough from each other that the inter-world influences are diminished.\footnote{This can be seen from the fact that the expression for $F_{Q_k}$ can be derived by discretizing the quantum potential in Eq. (\ref{Qpotential}), according to a standard nearest-neighbor finite-difference scheme. Thus, the Bohmian paths determined by Eq. (\ref{Euler3}) are recovered in the limit as the discretization step size approaches zero. See  App. \ref{app:CMIW-discretization} for details on the discretization process. See also \cite[App.~A]{hallQuantumPhenomenaModeled2014} for a detailed discussion of the connection between the quantum potential and the MIW inter-world potential.} We will use the term \emph{\(\mathcal{Q}\)-local} (configuration-space-local) to characterize inter-component influences that attenuate significantly when the inter-component distances, as measured using the natural metric for $\mathcal{Q}$, are large.\footnote{For the moment, we waive a precise specification of what constitutes ``significant attenuation'', although we presume that the proposed inverse-cube law will suffice. Thus, the particular discrete toy model of Hall et al. may be considered to be at least \(\mathcal{Q}\)-local.} This is in contrast to Bell's notion of \emph{local causality} (\cite[Ch.~24 ]{bellSpeakableUnspeakableQuantum2004}), which is strong enough to classify all action-at-distance influences as nonlocal. Local causality also differs from $\mathcal{Q}$-locality in that it is fundamentally a property of systems embedded in Minkowski spacetime, whereas $\mathcal{Q} $-locality is a property of systems located in $\mathcal{Q}$.

It is likewise useful to distinguish $\mathcal{Q}$-locality from a third notion of locality, one that characterizes the dynamical equations governing field variables on $\mathcal{Q}$. The Schr\"odinger equation and hydrodynamic equations exhibit what we will call
$\mathcal{N}$-\emph{locality} (neighborhood-locality): the time derivative of the fields ($\psi(\textbf{x},t)$, $\rho(\textbf{x})$, and $\textbf{v}({\textbf{x}})$) at
$\mathbf{x}\in\mathcal{Q}$ depends only on the value of the fields and a finite number of their derivatives in an
arbitrarily small neighborhood $\mathcal{N}_{\textbf{x}}$ of $\mathbf{x}$. Hence, the evolution of the fields at $\mathbf{x}$ does not
directly depend on data contained in distant regions of $\mathcal{Q}$ that do not overlap with $\mathcal{N}_{\textbf{x}}$.

In Sect. \ref{general} we will argue that $\mathcal{Q} $-locality is not peculiar to this 1D DMIW model, but is plausibly a necessary condition for any concrete realization of the DHV that reproduces the correct quantum behavior.

\subsection{Translating to the Standard Wavefunction Picture}\label{recover}

According to the DMIW framework, wavefunctions do not appear as fundamental elements of the theory. Since Eq. (\ref{WisemanTrajectory}) is second order in time, the fundamental state of the system at a particular time is given by a complete specification of the positions and velocities of each world. Nonetheless, it is possible to partially reconstruct the wavefunction once the DMIW state is specified.\footnote{The DMIW state space is in fact larger than the space of wavefunctions since some velocity fields are not expressible as a phase gradient. To ensure that every combination of $\rho(\textbf{x},t)$ and $\textbf{v}(\textbf{x},t)$ gives rise to a corresponding wavefunction an additional quantization condition: $\frac{1}{2\pi\hbar}\oint\sum_jm_j\vec{v}_j\cdot{\vec{dr}_j}\in\mathbb{Z}$ must be imposed (see \cite[Sect.~6]{sebensQuantumMechanicsClassical2015}, \cite{schiffCommunicationQuantumMechanics2012}, and \cite{reddigerMathematicalTheoryMadelung2023}). Note that CMIW \emph{does} preserve this condition, if satisfied at some initial time, over all times. It is not clear whether DMIW also does so, as the discrete nature renders it difficult to even define the circulation integral.} The modulus $R(\textbf{x},t)$ of the wavefunction is uniquely determined by the world density $\rho(\textbf{x},t)$ via Eq. (\ref{density}), and the phase $S(\textbf{x},t)$ of the wavefunction is determined by the velocity field up to an additive constant via Eq. (\ref{velocityfield}). Importantly, the translation between DMIW state data and the wavefunction is only possible when the worlds are packed tightly enough for $\rho(\textbf{x},t)$ to be meaningfully defined as a smooth approximation of the world density. Thus, from the DMIW perspective, the reason that we encounter phenomena that seem to be describable by wavefunctions is fundamentally due to the fact that the density of worlds is sufficiently high.  

The translatability of the DMIW and wavefunction descriptions of quantum systems provides a convenient way to connect the DMIW approach with the Everettian many-worlds interpretation, in which the wavefunction does play a central role. According to decoherence theory, a measurement-like interaction transforms an initial product state

    \begin{equation}\label{initial}
        \Psi_i(\mathbf{x},\mathbf{y})
        = \phi(\mathbf{y}) \sum_{j=1}^{J} C_j\,\psi_j(\mathbf{x}),
    \end{equation}

\noindent where $\phi(\mathbf{y})$ is the (initial or ``neutral'') joint detector--environment state and $\sum_j C_j \psi_j(\mathbf{x})$ is the state of the system to be measured, into the entangled superposition

    \begin{equation}\label{final}
        \Psi_f(\mathbf{x},\mathbf{y})
         = \sum_{j=1}^{J} C_j\,\phi_j(\mathbf{y})\,\psi_j(\mathbf{x}).
    \end{equation}

\noindent Here $\mathbf{x}\in\mathcal{Q}_X$ and $\mathbf{y}\in\mathcal{Q}_Y$ denote the configuration-space coordinates of the system and the detector--environment, respectively, and  $\phi_j(\mathbf{y})$ are the detector--environment states correlated with the system outcomes.

Under environmental decoherence, the \(\phi_j(\textbf{y})\) become approximately orthogonal and develop nearly disjoint support in $\mathcal{Q}_Y$, so that \(|\Psi_f(\textbf{x},\textbf{y})|^2\) separates into roughly nonoverlapping branch regions \(\Omega_j\subset\mathcal{Q}\) corresponding to \(C_j\,\phi_j(\textbf{y})\,\psi_j(\textbf{x})\).\footnote{We assume that interactions between the detector and the environment select the position basis as the preferred pointer basis, thereby localizing the detector states in position. This assumption is justified by the fact that a realistic interaction Hamiltonian $\hat{H}_{\text{int}}$ will involve spatially local couplings—i.e., $\hat{H}_{\text{int}}$ is constructed from functions of the position operator—which induces decoherence of the wavefunction in the position basis \cite{romanoDecoherenceBasedApproachClassical2023}.} Translating to the DMIW picture where \(\rho(\textbf{x},\textbf{y})\) in Eq. (\ref{density}) is the coarse-grained world density, the same branching process appears as a partitioning of worlds within $\mathcal{Q}$.\footnote{In the Everettian literature, the branches are often called \emph{worlds}. Here, however, we use \emph{branch} for a decohered component of the wavefunction (labeled ``$j$''), and reserve \emph{world} to denote a configuration space trajectory (labelled ``$k$''). In principle, each \emph{branch} can contain multiple \emph{worlds}.} As \(|\Psi(\textbf{x},\textbf{y})|^2\) divides into branches, the initial ensemble of $K$ total worlds divides into disjoint subensembles of $K_j$ worlds each (where``$j$'' labels a particular wavefunction branch). Thus, we end up with $K_j$ worlds in \(\Omega_j\), and with $\sum_{j=1}^J K_j =K$.
 At the coarse-grained level, the splitting of densely packed worlds into subensembles  occupying roughly disjoint regions of $\mathcal{Q}$ appears as a fluid dividing into branches.\footnote{In fact, the coarse-grained fluid picture may be what some Everettians have in mind when they visualize wavefunction branching. In contrast to branches, though, worlds are like cords or threads, which never intersect or divide. }

\subsection{Measurement Statistics}

 Recall that the motion of a single world $\textbf{x}_k$ in $\mathcal{Q}$ encodes the motion of $N$ particles in a three-dimensional space. This includes the motions of all the particles that make up the detectors and the physical records that constitute a particular measurement outcome. Measurement outcomes in a world $\textbf{x}_k$ are therefore grounded in the configuration of its particles. To recover the statistical predictions of quantum mechanics, we must show that outcomes resulting from a long-run sequence of measurements are expected to accord with quantum statistics in the DMIW framework. To this end, it will be illustrative to recount the difficulties the Everettian approach has in accounting for probabilities. 
 
 An initial worry one might have when first encountering the Everettian interpretation is how the Born rule is to be recovered. After a measurement-like interaction and decoherence, an initial state given by Eq. (\ref{initial}) evolves to a final state given by Eq. (\ref{final}). A natural first thought is to treat the decohered branches as a discrete set and assign equal weight to each branch (“direct branch counting”), which yields a probability $1/J$ per branch. This does not accord with the Born rule since in general $1/J\neq\lvert{C}_j\rvert^2$. In light of these difficulties, many Everettians have abandoned direct branch counting in favor of other strategies (notably decision-theoretic reconstructions \cite{wallaceEmergentMultiverseQuantum2012,deutschQuantumTheoryProbability1999}) that recover the Born rule at the expense of introducing additional axioms and epistemic principles.

In the DMIW scheme, in contrast, direct world counting works to ground probabilities, since the fraction of worlds that move into branch $\Omega_j$ at time $t$, given by 

    \begin{equation}
        \frac{K_j}{K} \;\approx\; \int_{\Omega_j}\!\rho(\textbf{x},t)\,d\mathbf x \;=\; |C_j|^2,
    \end{equation}

\noindent accords with the Born rule. This is enough to ground probabilities even on a single run of an experiment, since an observer initially unaware of which world she is in will expect to end up in a world occupying region $\Omega_j$ with a probability given by $\lvert{C_j}\rvert^2$---if she assigns an equal weight to each world.\footnote{If the experiment is well designed, then the regions $\Omega_j$ will sharply delineate particle configurations corresponding to distinct measurement outcomes $O_j$ so that $\mathbb{P}(O_j)\approx\mathbb{P}(\Omega_j)\approx\vert{C_j}\rvert^2$. } Direct world counting also reproduces long-run quantum statistics, since a large sequence of repeated measurements will generate a branching structure of world sub-ensembles such that typical worlds will exhibit a history with outcome statistics that reflect the Born rule. 

The DMIW approach also explains the curious duality of the wavefunction that appears in Bohmian mechanics. In principle, the probability distribution over possible locations of the Bohmian world may differ greatly in character from the guiding field that governs its motion; however, Bohmian mechanics takes both to be described by the wavefunction. To explain the dual nature of the wavefunction, Bohmians have either proposed that the two fields dynamically relax into alignment \cite{valentiniSignallocalityHiddenvariablesTheories2002} or they have argued that universes exhibiting Born statistics are typical according to a natural class of measures \cite{durrQuantumEquilibriumOrigin1992}. In the DMIW framework, the connection between probability and dynamics is straightforward since both are grounded in the behavior of the world ensemble. One cannot vary the probability distribution without altering the world density, which would in turn alter the dynamical behavior of the worlds.\footnote{That said, the precise form of the inter-world interaction that is presumed for $U$ must be chosen to be consistent with the Schr\"odinger equation in the dense-world limit, which does impose a constraint on the allowed forms for $U$.}

\subsection{Connection with the Broader DHV}
The concrete example provided by the DMIW framework above provides a template for how the more general DHV, as bare ontological commitment, can be completed into a full foundational theory. The first step is to determine a class of discrete components that constitute the Madelung fluid. They must be discrete beables localized within a sufficiently small region of $\mathcal{Q}$ so that their locations in $\mathcal{Q}$ correspond, at least approximately, to outcomes in a three-dimensional space.\footnote{Recall from Sect.~3.3 that a point $x \in \mathcal{Q}$ specifies the complete configuration of all $N$ particles in three-dimensional space, including the degrees of freedom associated with detectors and macroscopic records. Localization within a sufficiently small region of $\mathcal{Q}$ therefore corresponds to a definite macroscopic configuration in 3-space.} Once a class of discrete components is settled on, the DHV theorist can ground outcome probabilities in direct component counting, just as world-counting grounds the Born rule in the DMIW case. Finally, the DHV theorist must carefully choose her dynamics to guarantee that the ensemble of components will produce the correct long-run quantum statistics. In effect, the dynamics must yield the Madelung hydrodynamics in the high-density regime. Provided these elements are all in place, the DHV inherits the same interpretational resources as the DMIW framework, and thereby becomes just as capable of addressing the standard foundational issues. 

With these conceptual tools in place, we can now begin building toward the problem of sparse ontology. To that end, it will be useful to examine how the ideas developed in Sect. \ref{sec3} are realized in the concrete example of a Stern-Gerlach (SG) system, which plays a central role in the sparse-world argument developed in Sect. \ref{sparseworldargument}.  

    \begin{figure}
        \centering
        \includegraphics[width=0.5\textwidth]{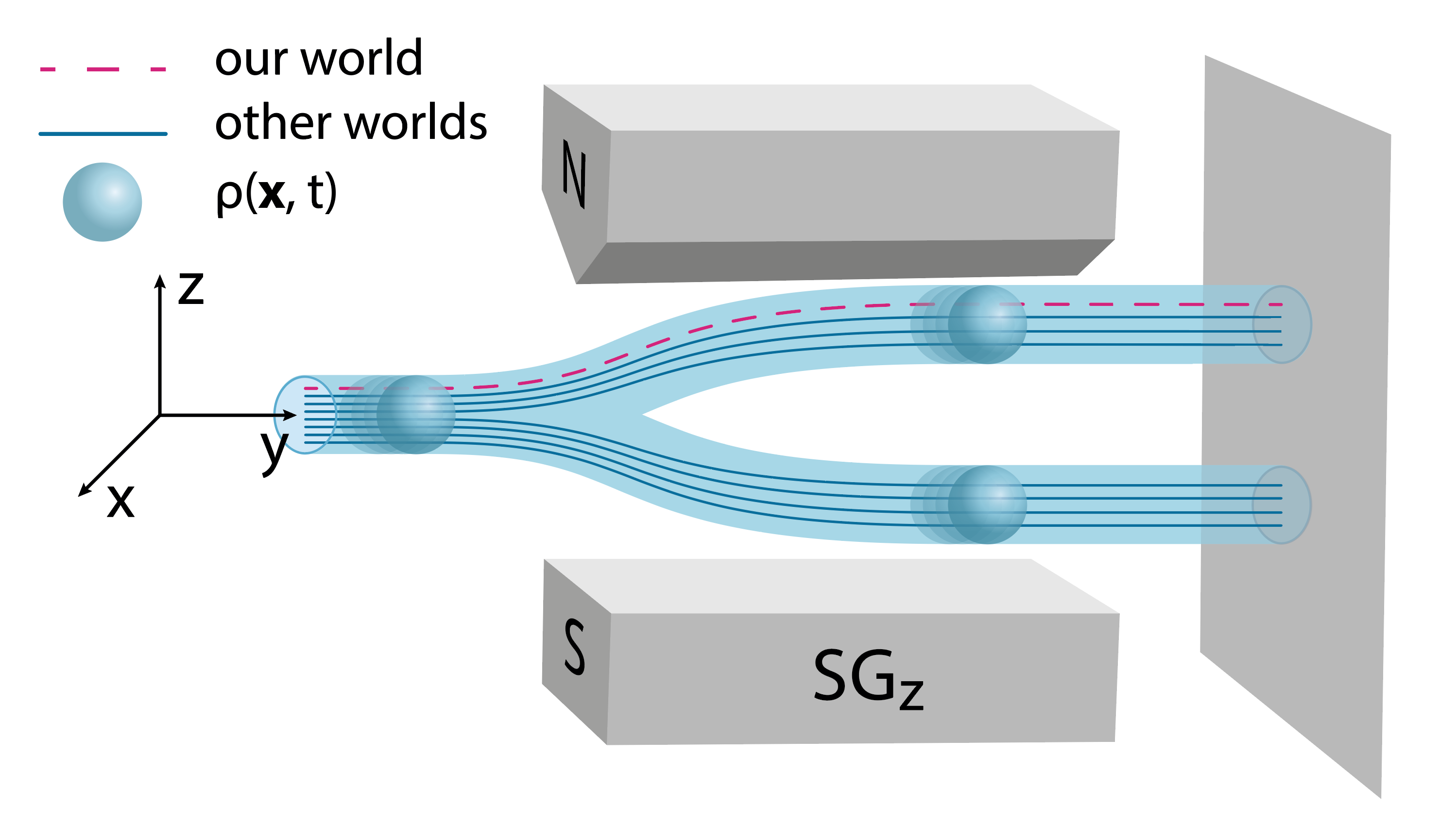}
        \caption{The hydrodynamic density, representing the Madelung fluid, is depicted by a blue spherical region to illustrate its localization around a small area (idealized here for clarity; in reality, the density is Gaussian). As the system passes through the analyzer, the density splits along the $z$-axis into two branches corresponding to the spin components. Overlaid are typical DMIW trajectories: the dashed pink line traces the path of our world, while the solid blue lines represent trajectories of other worlds in the ensemble. The worlds separate into two sub-ensembles displaced along $z$, mirroring the splitting of the hydrodynamic density.}\label{fig1}
    \end{figure}

\section{A Simple System that Creates Branch Splitting}\label{sec4}

Consider an SG system, where a neutral spin-$\frac{1}{2}$ atom\footnote{Here we have in mind the case of neutral silver atoms which interact with the magnetic field via an unpaired valence electron.} is incident on a SG analyzer that is oriented in the $z$-direction, and produces a magnetic field $\vec{B}=(0,0,bz)$.\footnote{The field $\vec{B}$ used here is unphysical since it doesn't satisfy the Maxwell conditions $\vec{\nabla}\cdot\vec{B}=0$ and $\vec{\nabla}\times\vec{B}=0$. Platt in \cite{plattModernAnalysisStern1992} justifies this common simplification by arguing that $\vec{B}$ is an approximation to the physical field $\vec{B}'=(bx,0,b_0-bz)$ (which does satisfy the Maxwell conditions) since rapid Larmor oscillations about the $z$-axis effectively decouple the system from the transverse inhomogeneity. Hsu et al. \cite[Fig.~7]{hsuSternGerlachDynamicsQuantum2011} show that, even with the transverse inhomogeneity included,  the wavefunction splits into roughly non-overlapping wave packets along the $z$-direction for a certain set of initial conditions. Since we are interested in studying the implications of this splitting, we use the simpler $\vec{B}$, which generates the same qualitative phenomenon.} To illustrate the relation between the standard wavefunction, CHV (e.g., with CMIW as a complete instantiation), and DMIW approaches, we will analyze the SG setup according to each approach in turn.

\subsection{The Wavefunction Approach}\label{wavefunction}
We take the initial state to be a two-component Pauli spinor propagating in the $y$-direction,  

    \begin{equation}
        \psi(\vec{x},0)=\phi(\vec{x},0)\begin{pmatrix}
               C_\uparrow\\
               C_\downarrow
            \end{pmatrix},
    \end{equation}

\noindent where $\phi(\vec{x},0)$ is a localized spherically symmetric Gaussian wave packet centered at the origin and $C_{\uparrow/\downarrow}\in\mathbb{C}$. The dynamics of the system are governed by the Pauli equation 

    \begin{equation}\label{Pauli}
        i\hbar\frac{\partial}{\partial{t}}\psi=-\frac{\hbar^2}{2m}\nabla^2\psi+\mu_B\vec{B}\cdot\vec{\sigma}\psi,
    \end{equation}

\noindent where $\mu_B=\frac{e\hbar}{2m_e}$ is the Bohr magneton.

Provided that the initial Gaussian wave packet is separable by Cartesian components, i.e. $\phi(\vec{x},0)=\phi_x(x,0)\phi_y(y,0)\phi_z(z,0)$, then for the magnetic field considered, the time evolution of Eq. (\ref{Pauli}) becomes similarly separable.  Thus, $\phi_x(x,t)$ and $\phi_y(y,t)$ evolve freely, while $\phi_z(z,t)$ splits into two distinct and (eventually) non-overlapping branches---i.e., $\phi_{z_\uparrow}(z,t)$ displaced along the positive $z$-direction, and $\phi_{z_\downarrow}(z,t)$ displaced along the negative $z$-direction. Therefore, for $t>0$, 

    \begin{equation}
        \psi(\vec{x},t)=\begin{pmatrix}
               C_\uparrow\phi_{\uparrow}(\vec{x},t)\\
               C_\downarrow\phi_{\downarrow}(\vec{x},t)
            \end{pmatrix}
    \end{equation}

\noindent where $\phi_{\uparrow}(\vec{x},t)$ and $\phi_{\downarrow}(\vec{x},t)$ are localized wave packets that separate from each other along $z$.  Since the initial packet $\phi(\vec{x},0)$ is taken to be spherically symmetric, each emerging branch $\phi_{\uparrow}(\vec{x},t)$ and $\phi_{\downarrow}(\vec{x},t)$ remains a Gaussian wave packet with the same time-dependent width in all three spatial directions. \footnote{Explicit expressions for $\phi_{\uparrow}(\vec{x},t)$ and $\phi_{\downarrow}(\vec{x},t)$ in natural units $\hbar=m=c=1$ for a slightly different field configuration $\vec{B}=(0,0,bz/e)$, are given in \cite[Eq.~21]{krekelsZigzagDynamicsStern2024}. From these expressions, one sees that the packets remain spherically symmetric at all times provided the initial widths satisfy $d_x=d_y=d_z$ (their notation).} Furthermore, we will generally take the initial wave packet to be a coherent Gaussian that is moving in the $y$ direction, but stationary in $x$ and $z$. Being initially coherent, each branch of the wave packet may be expected to disperse (spread out) in all three directions over time, due to the Laplacian term in Eq. (\ref{Pauli}). However, the system parameters may be chosen so as to keep dispersion to a minimum. This may become especially important, in practice, when many SG operations are performed sequentially (to be considered in Sect. \ref{sparseworldargument}).

Additionally, it is presumed that the time spent by the particle within the SG magnetic field suffices for $\phi_{\uparrow}(\vec{x},t)$ and $\phi_{\downarrow}(\vec{x},t)$ to have separated from each other in the $z$-direction by  a distance much larger than the branch wave packet width. This will ensure approximate non-overlapping support for the two branches, with total density given by

    \begin{align}
            \rho=\psi^\dagger\psi&\approx\lvert{C}_\uparrow\rvert^2\lvert\phi_{\uparrow}(\vec{x},t)\rvert^2+\lvert{C}_\downarrow\rvert^2\lvert\phi_{\downarrow}(\vec{x},t)\rvert^2,
            \\
            &=:\rho_{\uparrow}(\vec{x},t)+\rho_{\downarrow}(\vec{x},t)\label{splitdensity},
    \end{align}

\noindent where $\rho_{\uparrow}(\vec{x},t)$ and $\rho_{\downarrow}(\vec{x},t)$ are roughly non-overlapping local densities.\footnote{The standard convention here would have $\lvert{C}_\uparrow\rvert^2 + \lvert{C}_\downarrow\rvert^2 = 1$, so that
$\phi_{\uparrow}(\vec{x},t)$ and $\phi_{\downarrow}(\vec{x},t)$ are separately normalized to unity for all $t$. The values of $C_\uparrow$ and $C_\downarrow$ are thus completely determined apart from an immaterial global phase constant.} 
Crucially, if no further interaction mixes $\phi_{\uparrow}(\vec{x},t)$ and $\phi_{\downarrow}(\vec{x},t)$, the two packets thereafter evolve independently of each other. 

\subsection{The CHV Approach}\label{hydroapproach}

Following Sect. \ref{sec2}, we can translate from the wavefunction to hydrodynamic form by substituting the spinor polar decomposition\footnote{Here, the coordinate dependence of $\chi(\vec x, t)$ introduces a nontrivial ambiguity in the 
definition of $S(\vec x,t)$, corresponding to a gauge freedom.  Nevertheless, in the CMIW formalism, the resultant quantum trajectories turn out to be gauge-invariant---as they must be, if they are to have ontological significance. In any event, note that Eq. (\ref{chieq}) reflects one specific choice of gauge, in that the relative phase between spin-up and spin-down components is chosen to always satisfy a definite relationship.}

    \begin{equation}\label{spinpolar}
        \psi(\vec{x},t)=R(\vec{x},t)e^{iS(\vec{x},t)}\chi(\vec{x},t),
    \end{equation}

\noindent with $\chi^\dagger(\vec{x},t)\chi(\vec{x},t)=1$, into the Pauli equation (\ref{Pauli}). To recast everything in hydrodynamic variables, it is convenient to parametrize $\chi(\vec{x},t)$ in spherical coordinates as

    \begin{equation}
    \label{chieq}
        \chi(\vec{x},t) =
            \begin{pmatrix}
                \cos\!\left(\tfrac{\theta(\vec{x},t)}{2}\right)e^{i\phi(\vec{x},t)/2} \\
                i\,\sin\!\left(\tfrac{\theta(\vec{x},t)}{2}\right)e^{-i\phi(\vec{x},t)/2}
            \end{pmatrix}.
    \end{equation}

\noindent We then define the local spin field 

    \begin{equation}
        \vec{s}(\vec{x},t):=\frac{\hbar}{2}\left[\sin\theta(\vec{x},t)\sin\phi(\vec{x},t),\sin\theta(\vec{x},t)\cos\phi(\vec{x},t),\cos\theta(\vec{x},t)\right],
    \end{equation}

\noindent where the angle fields $\phi(\vec{x},t)$ and  $\theta(\vec{x},t)$, defined with respect to a fixed Euclidean frame, are interpreted as describing the local direction of the fluid's spin field.  

Since the initial wavefunction is separable, the hydrodynamic equations also separate into three uncoupled equations, one for each Cartesian spatial coordinate. The $z$-equations are given by 

    \begin{align}
        \dot{\rho}_z &= -\rho_z\partial_zv_z, \label{SGhydrocontinuity}\\[4pt]
        m\,\ddot{z} &= -\,\frac{\hbar^2}{4m\,\rho_z}\,\partial_z\!\left(\rho_z\,(\partial_z\phi)^2\sin^2\theta+\rho_z\,(\partial_z\theta)^2\right)
        - \partial_z Q_z + \mu_B\,b\,\cos\theta, \label{SGNSL}\\[4pt]
        \dot{\theta} &= \frac{\hbar}{2m\,\rho_z\sin\theta}\,\partial_z\!\left(\rho_z\,(\partial_z\phi)\,\sin^2\theta\right), \label{SGhydrotheta}\\[4pt]
        \dot{\phi} &= \frac{\hbar}{2m\,\rho_z\sin\theta}\Big[-\,\partial_z\!\big(\rho_z\,\partial_z\theta\big)
            + \rho_z\,(\partial_z\phi)^2\sin\theta\cos\theta\Big]
            + \frac{2\mu_B}{\hbar}\,b\,z, \label{SGhydrophi}
    \end{align}
 
\noindent where $Q_z$ is the one-dimensional single-particle quantum potential (we have suppressed the independent variables $\vec{x}$ and $t$).\footnote{See \cite[Ch.~9]{hollandQuantumTheoryMotion1993} and \cite{lombardiniInteractingQuantumTrajectories2024}
for detailed derivations of Eqs.~\eqref{SGhydrocontinuity}--\eqref{SGhydrophi}.} The $x$- and $y$-equations are the single-particle ($N=1$), free-case forms of Eqs. (\ref{continuity3}) and (\ref{Euler3}). Eqs. (\ref{SGhydrocontinuity}) and (\ref{SGNSL}) are the spin-adjusted analogs of Eqs. (\ref{continuity3}) and (\ref{Euler3}), describing, respectively, the conservation of fluid components and translational motion of the fluid along $z$.

In the CHV picture, the fluid is initially localized in a spherically symmetric high-density region centered at the origin as it enters the magnetic field, as illustrated by the blue spherical density profile in Fig.~\ref{fig1}. Passing through the magnetized region, it splits along the $z$-direction into two spherically symmetric high-density branches, with a fraction of its total density (roughly $\lvert{C}_\uparrow\rvert^2$) moving upward, and the remaining fraction ($\lvert{C}_\downarrow\rvert^2$) moving downward, according to Eq. (\ref{splitdensity}). The spin field $\vec{s}(\vec{x},t)$, which in general points away from $z$ throughout the fluid initially, rotates within the magnetized region such that it points predominantly in the positive $z$-direction in the upper branch, and in the negative $z$-direction in the lower branch.\footnote{See \cite[Fig. 9.13]{hollandQuantumTheoryMotion1993}.} Finally, once the branches are sufficiently well separated from each other, then the spin field within the upper branch points \emph{exclusively} in the $+z$-direction, whereas that in the lower branch points exclusively in the $-z$-direction. Thereafter, the hydrodynamic equations imply that each branch evolves essentially independently, since fluid elements only affect each other $\mathcal{N}$-locally. 

\subsection{The DMIW Approach}\label{DMIWapproach}

Finally, we 
consider the DHV picture, by specifying a concrete DMIW model for the SG system. Since the SG setup is inherently three-dimensional, the one-dimensional DMIW model discussed in Sect. \ref{DMIWdynamics} is not, by itself, sufficient to characterize the system. Fully multi-dimensional DMIW models have been proposed, but their dynamics are nontrivial\footnote{In one spatial dimension, the worlds can be globally ordered, so local density gradients and inter-world forces can be expressed directly in terms of nearest neighbors. In two or more dimensions, no such ordering exists. In \cite{herrmannEigenstatesManyInteracting2023}, Herrmann et al. use effective neighborhoods built around each world from Voronoi tessellations and Delaunay triangulations to estimate local density gradients and define the corresponding inter-world forces. This additional geometric step makes the dynamics substantially more complex and algorithmic in character.} and, to date, have not been developed to incorporate spin degrees of freedom in a systematic way.

To proceed, we instead consider a discretized version of the CMIW framework developed in \cite{lombardiniInteractingQuantumTrajectories2024} for single particles with spin. In this approach, one posits a continuum of world trajectories—one for each point in configuration space—labeled by coordinates $\mathbf{C} = (C_{1}, \ldots, C_{3N})$ where each value of $\mathbf{C}$ specifies a distinct world trajectory. To incorporate spin, each world is endowed with a spin vector, characterized by spherical angle variables $\theta(\mathbf{C}, t)$ and $\phi(\mathbf{C}, t)$. These spin variables play the same role as the spin-field variables introduced in the hydrodynamic description of Sect. \ref{hydroapproach}, but are now seen as being carried by individual worlds along their trajectories.

For initially separable wavefunctions, the corresponding CMIW framework is conveniently formulated using \emph{uniformizing coordinates} (See App. \ref{app:CMIW-discretization} for details.), which, when discretized, arranges the now-discrete ensemble of worlds onto a three-dimensional grid indexed by ($i,j,k$).\footnote{This is admittedly a very unlikely arrangement of worlds for a genuine DMIW model. Nonetheless, we consider this model because, as argued in Sect. \ref{general}, it exhibits the same qualitative behavior found in any plausible DMIW model.} If, in addition, the Schr\"odinger dynamics are separable, then the CMIW guidance equations also separate, thereby decoupling the motion of the worlds along each of the coordinate directions. In the particular case of the separable Pauli dynamics given by Eq. (\ref{Pauli}), the spin degrees of freedom couple only to the motion along the $z$-direction, implying that the spin variables $\theta$ and $\phi$ depend only on the $z$-label $k$. The equations of motion that govern the evolution of the shared $z$-coordinate and spin variables associated with each $k$-level of the world-grid are given by

    \begin{equation}\label{DMIWspinNSL}
        m\ddot{z}_k \;=\; F^z_{{Q_k}} \;+\; F^z_{{Q_{S_k}}}+\mu_Bb\cos\theta_k,
    \end{equation}

\noindent where 

    \begin{equation}\label{spinforce}
        \begin{aligned}
            F^z_{{Q_{S_k}}} \;=\; \frac{\hbar^2}{4m}\!\Bigg[
            &\frac{(\phi_k-\phi_{k-1})^2\,\sin^2\!\big(\tfrac{\theta_k+\theta_{k-1}}{2}\big)+(\theta_k-\theta_{k-1})^2}{(z_k-z_{k-1})^3}
            \\[-2pt]
            &-\;
            \frac{(\phi_{k+1}-\phi_k)^2\,\sin^2\!\big(\tfrac{\theta_{k+1}+\theta_k}{2}\big)+(\theta_{k+1}-\theta_k)^2}{(z_{k+1}-z_k)^3}
            \Bigg]
        \end{aligned}
    \end{equation}

\noindent is an additional inter-world force arising from spin interactions between the worlds.\footnote{In App. \ref{app:CMIW-discretization} we show that the equations of motion can be separated into three 1D expressions. For full derivations of Eqs. (\ref{DMIWspinNSL})-(\ref{SGDMIWPhi}), including the spin degrees of freedom, see \cite{lombardiniInteractingQuantumTrajectories2024}.}  The spin dynamics of each world are determined by 

    \begin{align}
        \dot{\theta}_k \;=\;& \frac{\hbar}{2m}\,\sin\theta_k\!
        \left[
        \frac{(\phi_{k+1}-\phi_k)\,\sin^2\!\big(\tfrac{\theta_{k+1}+\theta_k}{2}\big)}{(z_{k+1}-z_k)^2}
        -
        \frac{(\phi_k-\phi_{k-1})\,\sin^2\!\big(\tfrac{\theta_k+\theta_{k-1}}{2}\big)}{(z_k-z_{k-1})^2}
        \right], \\[4pt]
        \dot{\phi}_k \;=\;& \frac{\hbar}{2m}\!
        \Bigg[
        \left(\frac{\phi_{k+1}-\phi_{k-1}}{z_{k+1}-z_{k-1}}\right)^{\!2}\!\cos\theta_k
        -\frac{1}{\sin\theta_k}\!\left(
        \frac{\theta_{k+1}-\theta_k}{(z_{k+1}-z_k)^2}
        -\frac{\theta_k-\theta_{k-1}}{(z_k-z_{k-1})^2}
        \right)
        \Bigg]\nonumber
         \\
        &\;+\frac{2}{\hbar}\mu_B\,b\,z_k \label{SGDMIWPhi}.
    \end{align}

\noindent Here the $k$-levels are ordered along $z$, so that $k<k'$ implies $z_{k} < z_{k'}$. The $x$- and $y$-components remain purely translational and follow the free DMIW theory given by Eqs. (\ref{DMIWNSL}) and (\ref{DiscreteTrajectory}).

In this \emph{grid-based} DMIW model, the ensemble of $K$ worlds is initially packed into a localized region near the origin as it enters the magnetic field, as illustrated by the world trajectories in Fig.~\ref{fig1}. As it traverses the magnetized region, the Zeeman interaction term in Eq. (\ref{DMIWspinNSL}) drives the $k$-levels of the grid to split with $K_\uparrow \approx \lvert{C}_\uparrow\rvert^2 K$ of the worlds shifting up and $K_\downarrow \approx \lvert{C}_\downarrow\rvert^2 K$ shifting down---separating the ensemble into two dense sub-ensembles that constitute the hydrodynamic branches.\footnote{See \cite[Fig. 6]{lombardiniInteractingQuantumTrajectories2024}} The spin vector $\vec{s}_k$ of each world, which may initially be directed away from the $z$ axis, rotates to point in the positive/negative $z$-direction, for worlds that end up in the upper/lower branch.\footnote{See \cite[Fig. 6. a]{lombardiniInteractingQuantumTrajectories2024}. Interestingly enough, $\theta_k(t)$ can become, say, larger than $\pi/2$ during the course of its time evolution, before ultimately changing $z$-direction and converging to $\theta_k(t\rightarrow \infty)=0$.} 

It is important to note a peculiar feature of this model, one that we will later argue is inherited by any plausible DMIW model. In the wavefunction and CHV (including CMIW) approaches, the densities of the up and down branches, $\rho_{\uparrow/\downarrow}(\textbf{x},t)$, are spherically symmetric about their respective centers. In order for the up and down sub-ensembles to reflect this symmetry, neighboring $k$-levels of the grid must separate from each other so that the average inter-world distance along the $z$-direction is twice as large as the average inter-world separation along the $x-$ and $y-$ directions. This is because only the $k$-levels of the initial branch split into the up and down branches while the $i$- and $j$-levels do not. In short, the world density thins in a non-isotropic manner along the direction of the SG splitting. 

As in the wavefunction and hydrodynamic approaches, the evolution of the two final subensembles is thereafter independent. Here, however, this property arises from the $\mathcal{Q} $-locality of the inter-world interactions, Eqs. (\ref{spinforce})--(\ref{SGDMIWPhi}), whose influence dissipates once the worlds become significantly separated from each other.\footnote{In particular, we envision macroscopic final separations between the two sub-ensembles, e.g., on the order of mm's or larger. This also allows for quite large subensemble regions to be used, which automatically ensures that dispersion is minimal. Both features are highly relevant for the sparse world argument that follows.}

\section{The Sparse World Argument}\label{sparseworldargument}
\begin{figure}
        \centering
        \includegraphics[width=0.9\textwidth]{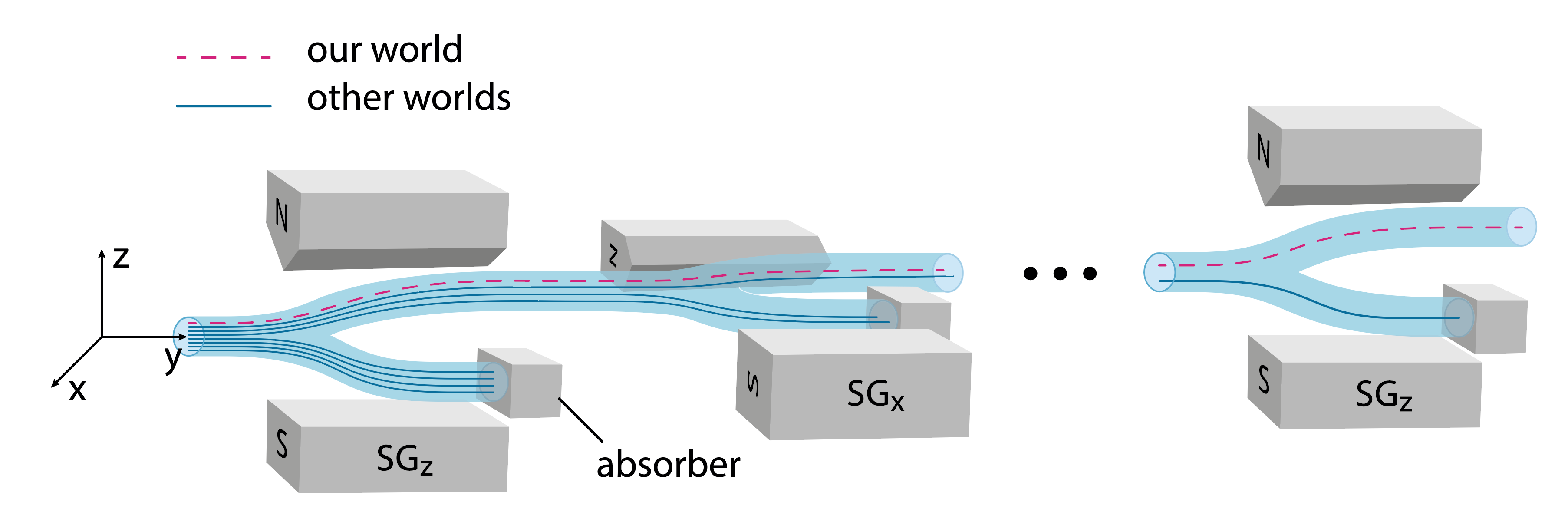}
        \caption{Schematic of the concatenated SG sequence used to illustrate repeated branching. The dashed pink line indicates the trajectory of our world, while the solid blue lines trace the trajectories of other worlds in the discrete world ensemble. The blue tube depicts the high-density region of the hydrodynamic distribution $\rho(\vec{x},t)=\lvert\psi(\vec{x},t)\rvert^2$ from standard quantum theory, which branches over time, but remains continuous (in spacetime). As branching proceeds, discrete trajectories diverge and become sparse relative to this persistent density, illustrating the onset of sparse ontology.}\label{fig2}
    \end{figure}

So far, the DHV---realized as a toy  DMIW model---accounts well for the basic SG system. Difficulties arise, however, when the system is modified. Consider a long sequence of SG analyzers, alternating between the $z$- and $x$-orientations. In each stage, the up-packet is allowed to pass through, while the down-packet is captured by an absorber (see Fig.~\ref{fig2}). We will argue that, under such concatenations, the DMIW predictions must inevitably deviate from those of the standard wavefunction and CHV/CMIW approaches, in an unacceptably major fashion, due to ``sparsening'' of the world density. 

To display the qualitative features of the sparsening mechanism, we work with the same grid-based DMIW model introduced in Sect. \ref{DMIWapproach}, while only considering the degrees of freedom of the particle---ignoring the degrees of freedom of the SG apparatus, absorber, and all other particles---so that $\mathcal{Q}$ is effectively $\mathbb{R}^{3}$. In Sect. \ref{general}, we argue that the same sparsening phenomenon arises even when realistic features are reintroduced.

Since the down-packet is effectively removed by the absorber,\footnote{In the wavefunction picture, interactions with the absorber induce environmental decoherence on the down-packet---thus greatly increasing the $\mathcal{Q}$-space separation from the up-packet still further, and ensuring the down-packet's subsequent dynamical isolation.} the spinor entering the $(j+1)$th SG analyzer can be taken to be the normalized up-component of the spinor exiting the $j$th SG analyzer. If each stage has transit time $T$, then this is 

    \begin{align}
        \psi_{j+1}(\vec{x}) = \psi(\vec{x},jT)&=\begin{pmatrix}
               \phi_{j_{\uparrow}}(\vec{x},jT)\\
               0
            \end{pmatrix}_{z/x},
            \\
            &= \frac{1}{\sqrt{2}}\phi_{j_{\uparrow}}(\vec{x},jT)\begin{pmatrix}
               1\\
               1
            \end{pmatrix}_{x/z},
    \end{align}
    
\noindent where $z/x$ denotes the basis of the spinor representation. Note that the width of each wave packet branch that exits a given SG apparatus is roughly the same as the width of the corresponding incoming packet, provided that dispersion is inconsequential \cite{lombardiniInteractingQuantumTrajectories2024}. Importantly, the description of each SG stage in terms of spinor wavefunctions is unaltered, and the sequence could, in principle, proceed indefinitely without requiring any change in its physical characterization. A similar story can be told in the CHV/CMIW picture without much difficulty, since the fluid simply splits into two branches upon each stage, with the lower branch always flowing into the absorber. However, things are quite different when we move to the DMIW picture.  

In the DMIW framework, the world ensemble acts fluid-like at the start of the SG sequence, but this behavior cannot be maintained for indefinitely long sequences. 
During each stage, roughly half of the worlds in the initial distribution shift up, while half are captured by the absorber. Then we can immediately see that the fluid approximation must eventually break down, since in the limiting case, a collection of $K$ worlds entering the SG sequence will be diluted \emph{exponentially quickly} down to a single world, via successive division.

In fact, the fluid approximation will likely break down long before $K_j \rightarrow 1$, since the smooth density $\rho(\vec{x},t)$ and velocity field $\vec{v}(\vec{x},t)$ will become under-resolved as soon as the worlds constituting each branch become sparse. Moreover, the breakdown of the fluid approximation is accompanied by a dynamical difficulty. Not only is the number of worlds reduced in half as the worlds pass from one SG analyzer to another, but the average distance between each level of the world grid is also effectively doubled along the direction of splitting (i.e., the $i$-levels for the $x$-oriented analyzer and the $k$-levels for the $z$-oriented analyzer), since we have argued that the widths of the incoming and outgoing wave packets are roughly the same. The trajectory of each world, as determined by Eqs. (\ref{DMIWspinNSL})--(\ref{SGDMIWPhi}), then gradually deviates away from the corresponding Bohmian path and starts to behave more classically in its $x$ and $z$ components, since the dynamics of this model are separable and the influence of neighboring worlds along those directions is diminished. This implies that DMIW predictions will differ noticeably from the predictions of quantum mechanics well before the worlds have passed through a sufficient number of SG analyzers to reduce $K_j$ to one. 

\begin{figure}[htbp]
  \centering
  \begin{minipage}[t]{0.45\linewidth}
    \centering
    \includegraphics[width=\linewidth]{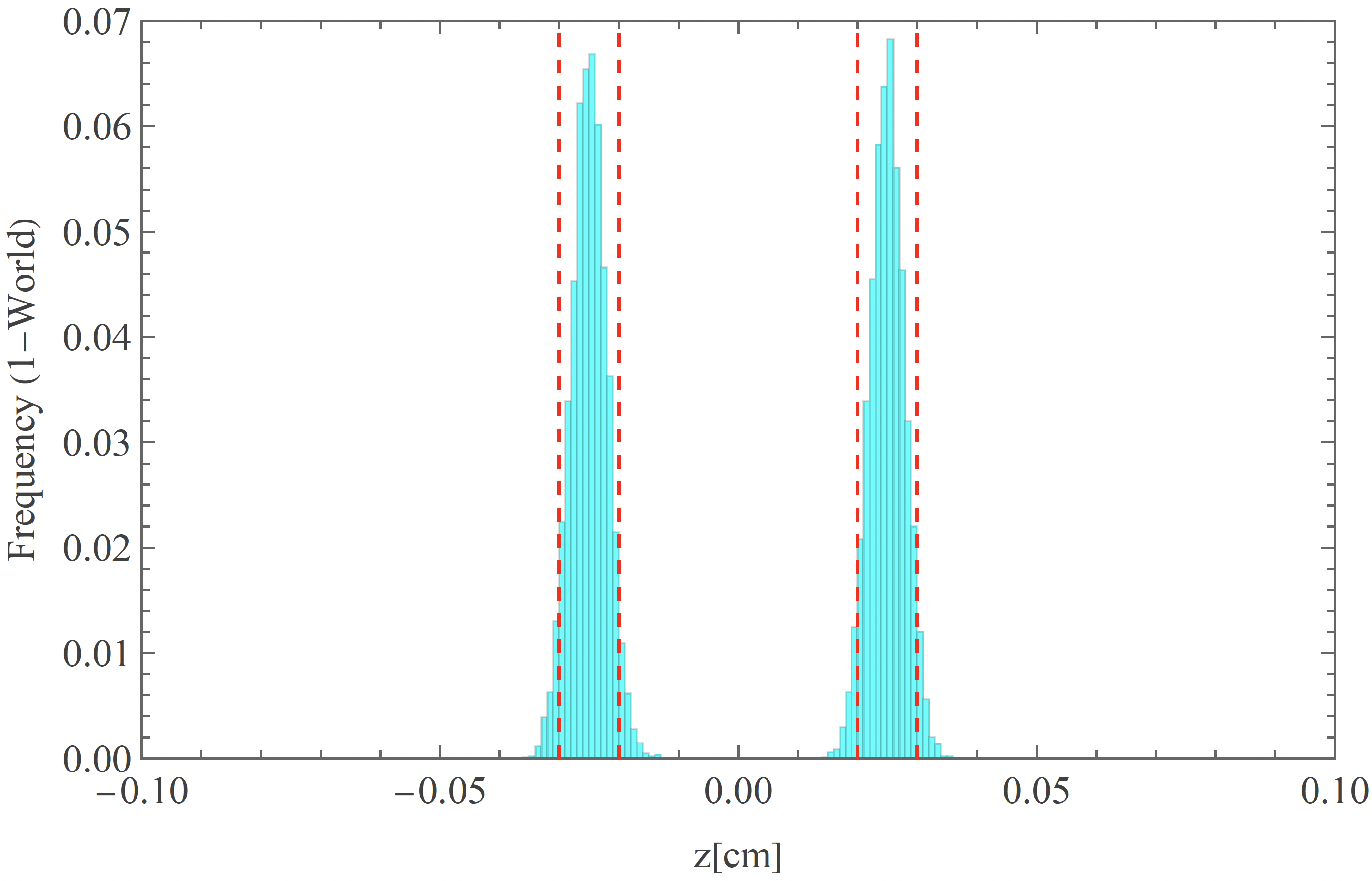}\\[-4pt]
    \text{(a)}
  \end{minipage}\hfill
  \begin{minipage}[t]{0.45\linewidth}
    \centering
    \includegraphics[width=\linewidth]{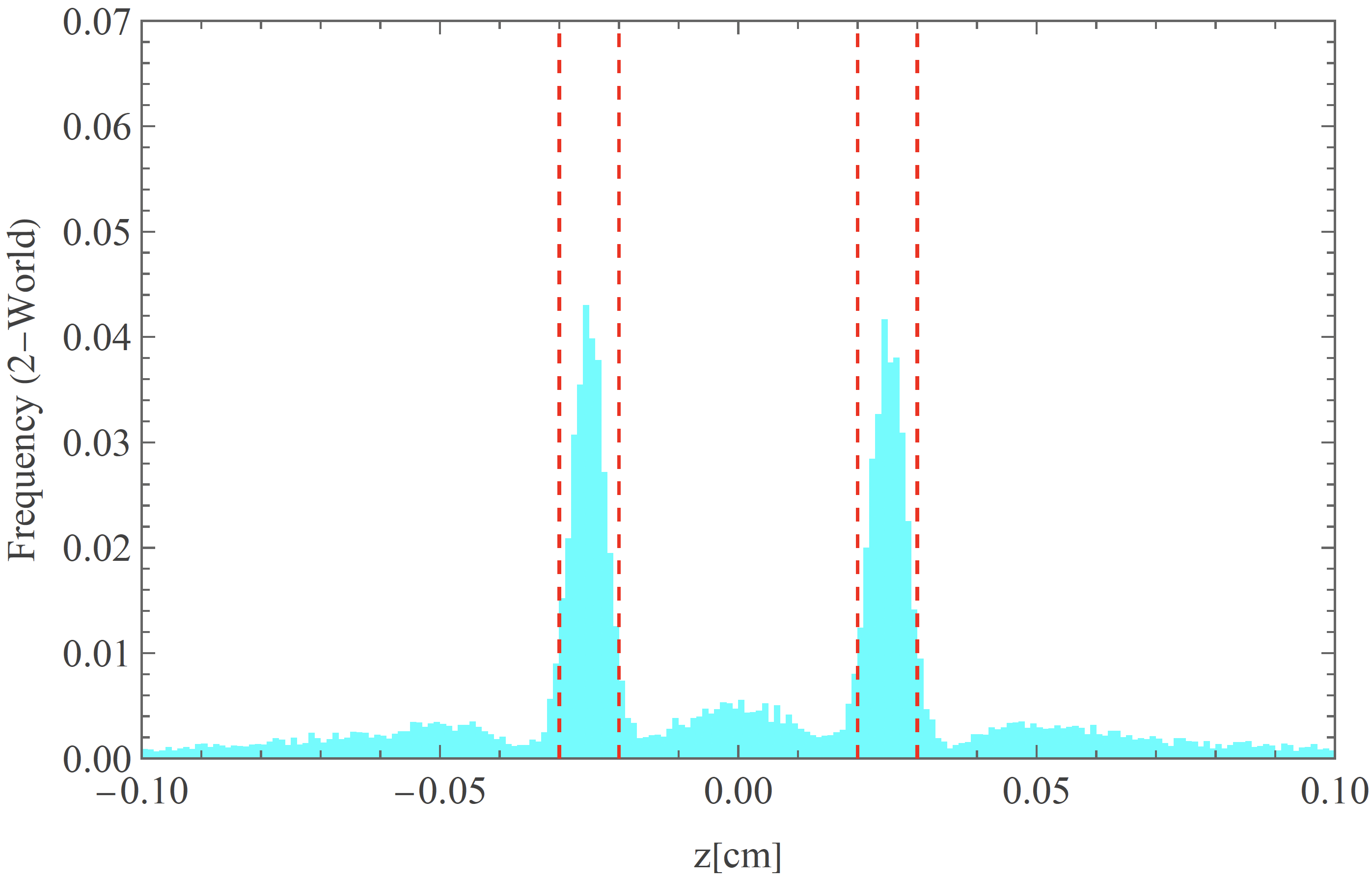}\\[-4pt]
    \text{(b)}
  \end{minipage}

  \medskip

  \begin{minipage}[t]{0.45\linewidth}
    \centering
    \includegraphics[width=\linewidth]{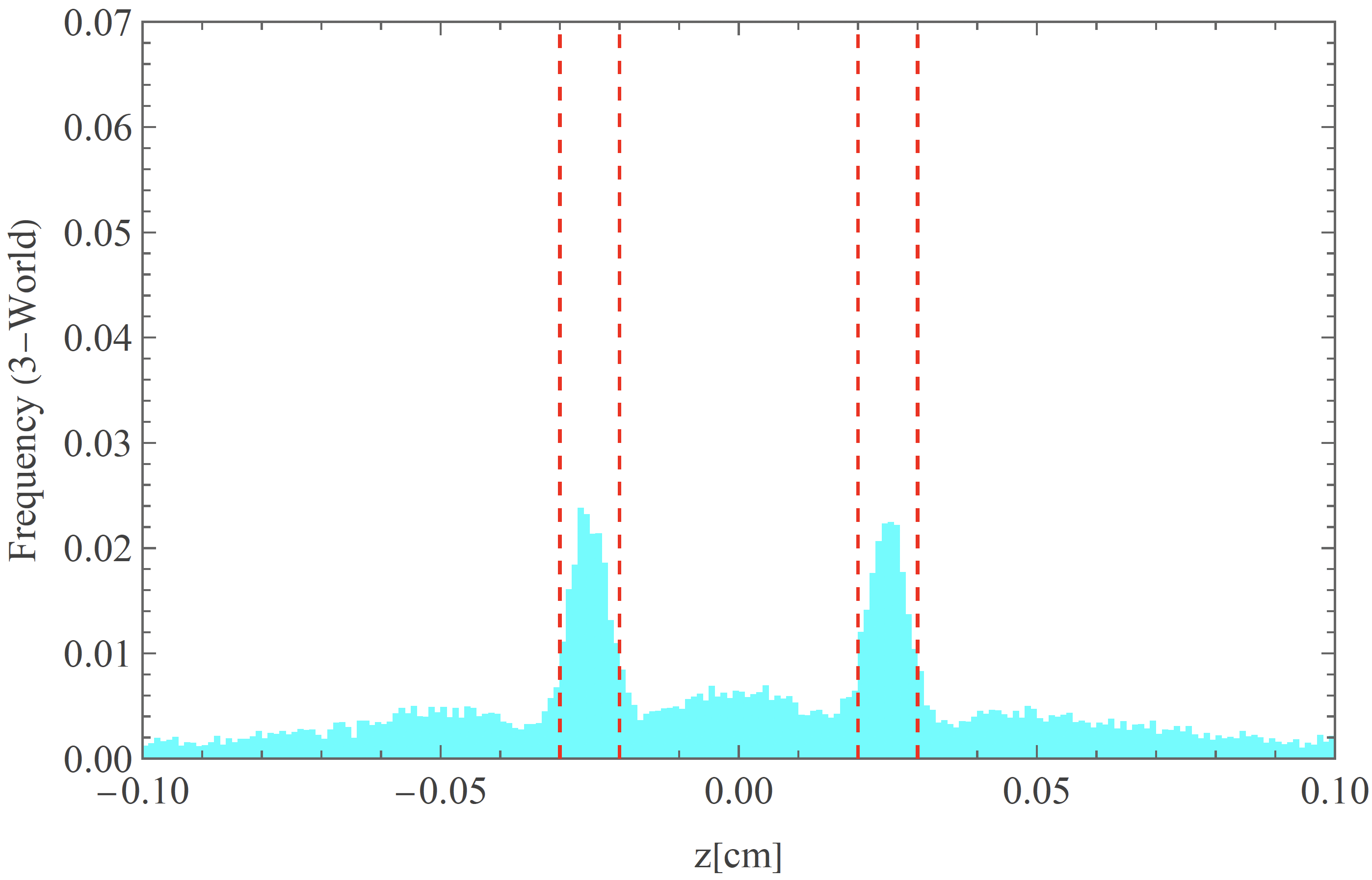}\\[-4pt]
    \text{(c)}
  \end{minipage}\hfill
  \begin{minipage}[t]{0.45\linewidth}
    \centering
    \includegraphics[width=\linewidth]{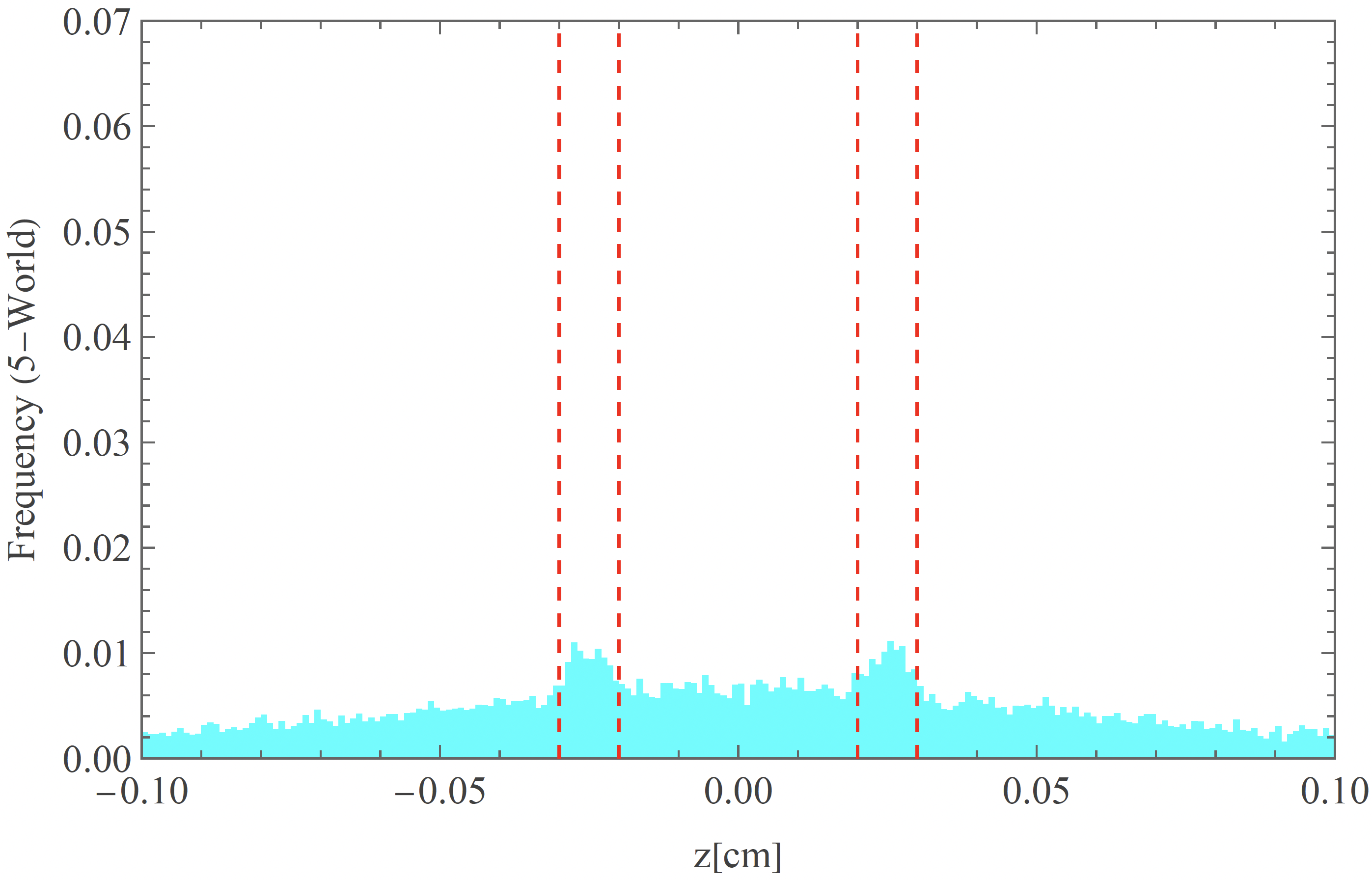}\\[-4pt]
    \text{(d)}
  \end{minipage}

  \medskip

  \begin{minipage}[t]{0.45\linewidth}
    \centering
    \includegraphics[width=\linewidth]{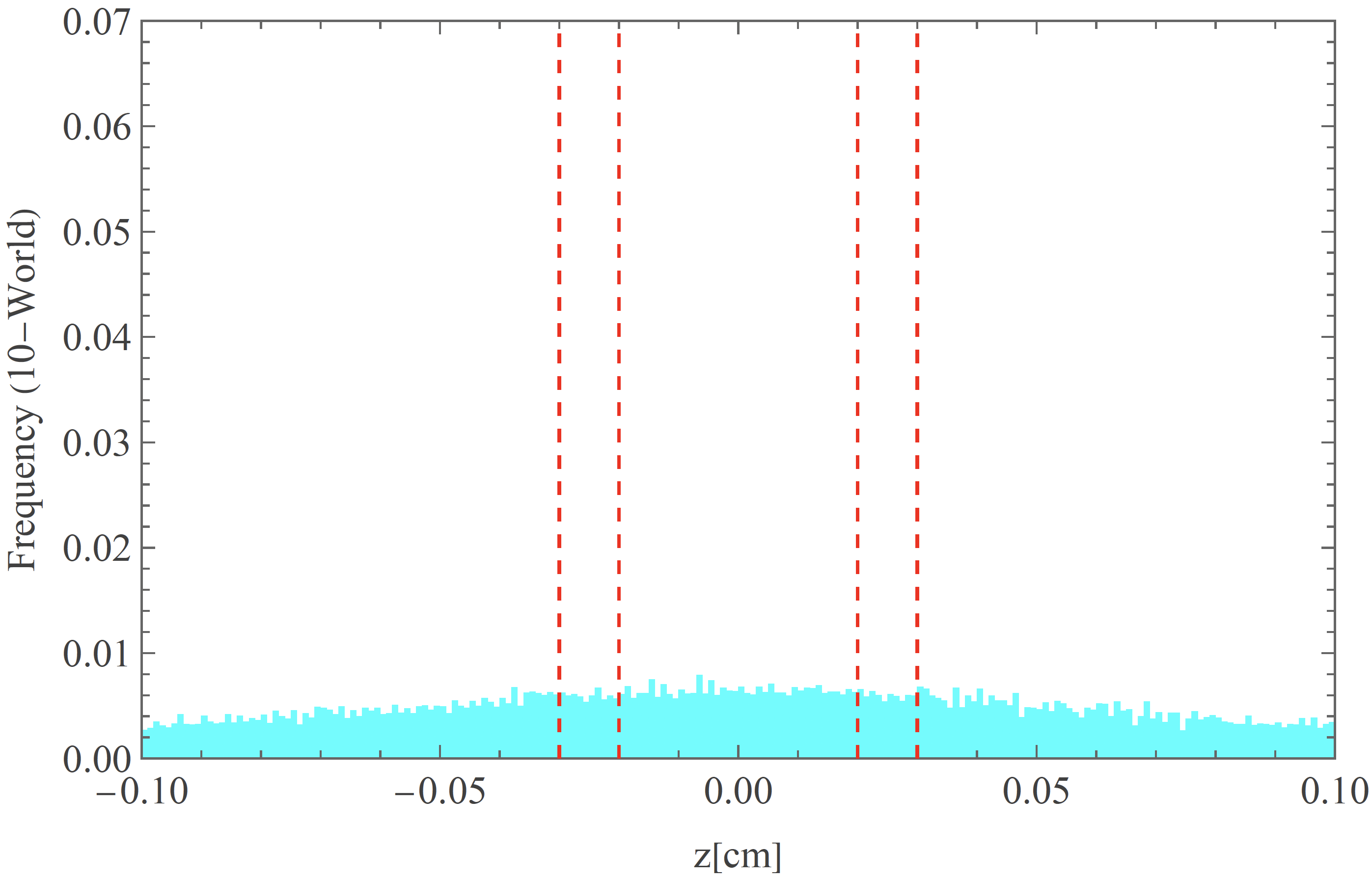}\\[-4pt]
    \text{(e)}
  \end{minipage}\hfill
  \begin{minipage}[t]{0.45\linewidth}
    \centering
    \includegraphics[width=\linewidth]{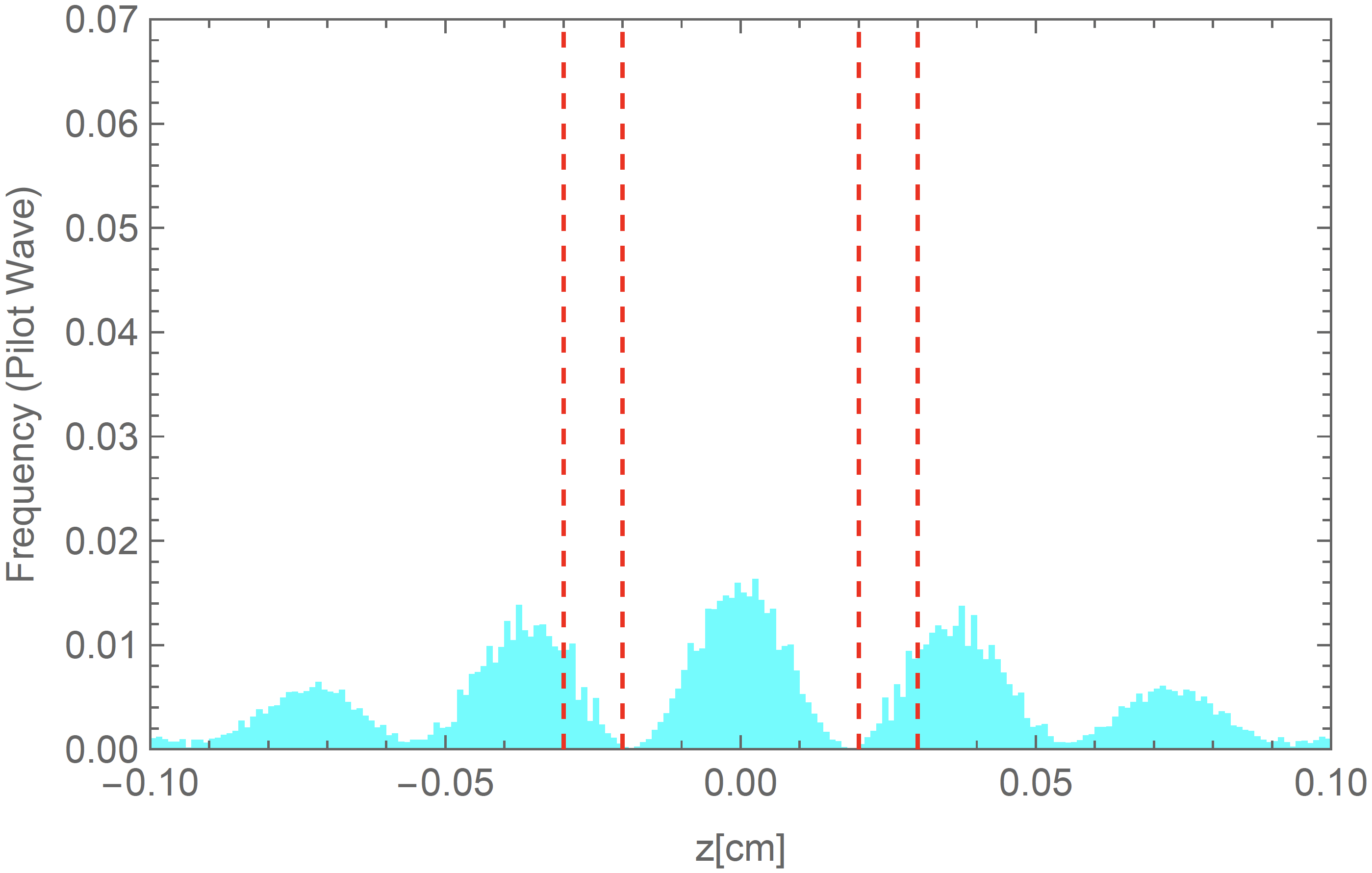}\\[-4pt]
    \text{(f)}
  \end{minipage}

  \caption{Normalized histogram plots of the interference patterns resulting from $2\times10^4$ runs for (a) 1, (b) 2, (c) 3, (d) 5, and (e) 10 DMIW worlds
   passing through a double-slit barrier of slit width 0.01 mm (dotted red ) at a time. Panel (f) shows the standard pilot-wave prediction, consistent with quantum mechanics.}
  \label{DoubleSlit}
\end{figure}

It is instructive to gain a better sense of exactly how the measured particle starts to lose 
its ``quantum character'' over time. To test for deviations from quantum mechanics, the above SG sequence can be interrupted at any stage, by directing the particle towards a double-slit interferometer, as suggested above. One can simulate the results that would be obtained from the above procedure by simply performing the DMIW double-slit calculation with a decreasing number of worlds.

Fig. \ref{DoubleSlit} shows the resulting hit distributions on the detection screen when only a few worlds---evolving under the free dynamics of Eq. (\ref{DiscreteTrajectory}) for a silver atom with $m=196,632\,m_e$---pass through a double-slit, of slit-width 0.01mm and slit separation 0.05mm, per run. The initial positions of the worlds are sampled from a bimodal Gaussian distribution (one per slit) of width $\sigma\approx.03$mm, and the transit time to the screen is taken to be 0.5 $\mathrm{s}$. In the limiting case where only a single world emerges from the SG sequence per run [panel (a)], the final distribution simply reproduces the initial distribution, since an isolated world moves inertially in the absence of nearby worlds. When two, three, or five worlds exit simultaneously [panels (b),(c), and (d), respectively], the final distributions show clustering behind each slit but are flattened relative to the initial distribution, since strong short-range inter-world repulsion prevents the worlds from arriving too close together on the screen. In the case of ten worlds [panel (e)], clustering behind each slit disappears, and the distribution becomes nearly flat with a slight bulge centered at $z=0$. By contrast, the prediction from standard quantum theory (and from Bohmian pilot-wave theory) is shown in panel~(f), which exhibits the familiar interference fringes. The sparse-world cases (a)--(e) fail to reproduce those fringes.

The concatenated SG sequence introduced above is intended as a proof-of-concept thought experiment, designed to illustrate how repeated branching leads to severe dilution of discrete hydrodynamic components in configuration space. While the qualitative conclusions drawn from this construction are robust, the idealized setup described here is not directly realizable in practice, due in particular to the rapid reduction of particle flux under successive absorptive measurements. A more realistic experimental variant that avoids this flux diminution problem—while preserving the essential features needed to probe the sparse-ontology effect—is outlined in App. \ref{secB}.

\section{Generalizing the Argument}\label{general}

It may seem that the sparse-ontology problem is a peculiarity of the \emph{grid-based} DMIW model introduced in Sect. \ref{DMIWapproach}, and that more realistic variants could evade it altogether. However, the features of the model that generate this problem are precisely those that any plausible DMIW formulation is likely to possess.\footnote{Indeed, the model was, in a sense, contrived to give DMIW the \emph{best shot possible} at avoiding sparse ontology difficulties (e.g., through its use of unrealistic rectilinear grids).} In the SG sequence thought experiment, deviations from standard quantum mechanics arise because (i) the average inter-world separation increases along the direction of the SG-induced branch-splitting, and (ii) the inter-world interactions responsible for producing quantum effects attenuate, since they are $\mathcal{Q} $-local. Can either of these features be plausibly avoided? Let us examine each in a bit more detail.

\subsection{Why World Thinning Is Unavoidable}
We now examine three families of proposals one might consider for avoiding world thinning: (i) forcing the world ensemble not to split, (ii) allowing world creation, and (iii) postulating microscopic redistribution dynamics.

\subsubsection{Non-Splitting World Dynamics}
In the grid-based DMIW model, the average inter-world separation along the splitting direction increases as the worlds pass through the SG sequence, since each branch inherits roughly the same spatial support in that direction as the parent packet while containing only half of the worlds. Thus, to prevent significant world separation, one might consider a DMIW variant in which, upon entering each SG analyzer, the entire flow of worlds follows a single branch—either all up or all down—thereby preserving the density of the initial ensemble. 

This proposal, however, is incompatible with the basic structure of the DMIW framework for at least two reasons. First, the Madelung hydrodynamic equations, to which a DMIW  model is supposed to converge in the high-density limit, describe a fluid that does split under SG evolution, since they are equivalent to the underlying Pauli dynamics. If the world density is to approximate the hydrodynamic density, then the world ensemble must split whenever the Madelung fluid splits. Forcing all worlds into a single branch breaks that correspondence.

Second, even if one were willing to abandon Madelung hydrodynamics for ``non-splitting" dynamics, this would conflict strongly with the probability interpretation championed by DMIW supporters, since probabilities would no longer be grounded in direct world counting. If a spin measurement were to be performed at the end of any given SG stage, the model should assign a probability of $1/2$ for ``our'' world to end up in the up-branch, and 1/2 for the down-branch. Direct world counting easily achieves this if the world sub-ensembles split, since one half of the worlds move up or down. In the ``non-splitting'' picture, however, this is impossible since 100\% of the worlds move along a single branch. 

More generally, the DMIW probability interpretation relies on the world density obeying the same continuity equation as the Madelung hydrodynamic density. If the worlds failed to split where the Madelung fluid splits, the world density would no longer track $\lvert\psi\rvert^2$, and direct world counting would fail to reproduce the Born rule. 

Thus, non-splitting dynamics should be rejected unless one is willing to abandon direct world counting altogether. 

\subsubsection{World Creation}
 The above discussion suggests another strategy to evade world-thinning. What if we postulate that worlds are created as the fluid splits, so that the density of worlds per branch does not decrease? The hydrodynamic continuity equation \eqref{continuity3} expresses the $\mathcal{N}$-local conservation of the Madelung fluid. In the DHV, this equation is meant to emerge from the underlying world dynamics in the high-density regime. At the coarse-grained level, creation or annihilation of worlds corresponds to adding a source term $\mathcal{S}(\mathbf{x},t)$ to the continuity equation,

    \begin{equation}
        \frac{\partial\rho(\vec{x},t)}{\partial t}
         = -\sum_{i=1}^N \vec{\nabla}_i\cdot(\rho(\textbf{x},t) \vec{v}_i(\textbf{x},t)) + \mathcal{S}(\textbf{x},t),
    \end{equation}

\noindent which is already inconsistent with the unitary Schr\"odinger evolution from which the hydrodynamic equations were derived. Nonetheless, one may be justified in introducing such source terms if the resulting dynamics reproduce the correct empirical statistics.

The difficulty is that the creation process would have to be finely adjusted to preserve the Born statistics under arbitrary branching patterns. For example, in an SG experiment, one would have to stipulate that whenever a world ensemble splits, new worlds are created in each sub-ensemble in such a way that: (i) the number of worlds per branch remains high enough to sustain the hydrodynamic approximation, and (ii) the ratio of world numbers in different sub-branches continues to track $|C_j|^2$ for the corresponding branch amplitudes. In effect, the creation rule must ``know'' the very quantum statistics that the DHV is supposed to reproduce from purely microphysical considerations. This threatens to reintroduce the wavefunction (or an equivalent informational structure) as a hidden, law-like ingredient that governs world creation, thereby undermining the original motivation to replace the wavefunction with a discrete hydrodynamic ontology.\footnote{This is reminiscent of the Lorentz-Fitzgerald attempts to explain away the lack of experimental observations of the ``ether drift''. Ultimately, the simpler explanation becomes much more palatable.} Thus, it seems plausible that the total world number is conserved.

\subsubsection{World Redistribution}

Finally, one might consider dynamics in which the worlds redistribute themselves evenly within each branch during the SG splitting process.\footnote{Strictly speaking, a complete even redistribution would destroy the Gaussian profile of the packets. What we have in mind instead is a local evening effect, in which the worlds rearrange themselves only locally so as to remove any inhomogeneity in their distribution. This could, perhaps, be brought about through a particular form of interworld potential that (energetically speaking) severely penalizes world distributions that are less even.} Such homogenizing dynamics could, in principle, make the increase in inter-world separation negligible, since although the world sub-ensembles constituting each branch are necessarily reduced in number during an SG splitting event, uniformly redistributing the remaining worlds within each branch would spread the reduced world number across all coordinate directions rather than primarily along the single splitting direction. For the unrealistic grid-based DMIW model formulated on a three-dimensional configuration space, redistribution is insignificant, but for a realistic DMIW theory in which the world sub-ensembles are distributed on the full $3N$-dimensional configuration space and are not restricted to lie on a grid in the subregion of the SG particle's configuration space, the effect is profound.\footnote{To appreciate this, consider a $3N$-dimensional branch region filled with worlds of minimum inter-world distance $l$. In the idealized case where the worlds are arranged on a uniform lattice (one per cell), the density is $\rho=1/l^{3N}$. After $K$ SG stages, the number of worlds in the branch is reduced by a factor of $1/2^K$, so uniform redistribution would give a new density $\rho_K=\rho/2^K=1/l_K^{3N}$. Thus $l_k/l=2^{K/3N}$. For a realistic universe-sized configuration space ($N\sim 10^{80}$), the exponent $K/(3N)$ is astronomically small for any $K$ realizable in the lab, making the post-redistribution change in inter-world separation vanishingly small.}     

However, such redistribution dynamics are unlikely to be characteristic of any realistic DMIW. For one thing, the streamlines of the hydrodynamic equations are the same as the Bohmian trajectories, which do separate from each other along the SG splitting direction but not significantly along the other coordinate directions. Since the hydrodynamic streamlines describe the average motion of the microscopic components of the fluid (in this case, worlds), the corresponding DMIW dynamics should generate approximately the same behavior in the high-density regime. 

If one is willing to consider modifications of the trajectory dynamics away from the presumed Bohmian form, then there is some freedom to modify the streamlines in a way that maintains probability conservation. This is because the hydrodynamic velocity is determined only up to the addition of a curl term,

    \begin{equation}
        \vec{v}_i(\textbf{x},t)\to\vec{v}_i(\textbf{x},t)+\frac{1}{\rho(\textbf{x},t)}\vec{\nabla}_i\times\vec{A}(\textbf{x},t), 
    \end{equation}   

\noindent which leaves the continuity equation unchanged. However, modifications that yield the correct redistribution of worlds are unlikely to be simple. The difficulty is that the Bohmian streamlines thin worlds along the splitting direction, but not along other orthogonal directions. This generates an anisotropy in the distribution of worlds that the hydrodynamic density---as a coarse-grained average---is insensitive to (i.e., $\rho(\textbf{x},t)$ is a scalar). Any curl term added to the original Bohmian velocity field must therefore be sensitive to these anisotropies. This makes it unlikely that a simple density-based expression could generate the desired redistribution dynamics.\footnote{There is also good reason to think the curl correction to the Pauli current is uniquely determined by the nonrelativistic limit of the Dirac equation \cite{hollandUniquenessConservedCurrents2003}. This modification doesn't result in uniform redistribution dynamics.}  

There is also a more general reason to think that worlds do not redistribute themselves evenly at the microscopic level. If they did endeavor to locally rearrange themselves in a homogeneous distribution, then we would expect to see this behavior manifest itself in the hydrodynamic regime through a Navier-Stokes-like pressure-gradient term,  

    \begin{equation}
        m_j\frac{d{\vec{v}(\textbf{x},t)}}{dt}=-\vec{\nabla}_jP(\textbf{x},t)+\ldots,
    \end{equation}

\noindent where $P(\textbf{x},t)$ is a simple function of $\rho(\textbf{x},t)$, whose effect would be to flatten $\rho(\textbf{x},t)$ toward a uniform distribution. In contrast, the Madelung equations feature the quantum force term $-\vec{\nabla}_jQ(\textbf{x},t)$, which does not drive the system into an equilibrated state. Instead, it supports persistent interference patterns (over time scales far exceeding the transit time through an SG analyzer) as well as stationary states with highly nonuniform structure. The fact that a high-density DMIW limit reproduces Madelung hydrodynamics, rather than classical Navier–Stokes hydrodynamics, therefore makes redistribution dynamics highly implausible.

\subsubsection{World Thinning: Summary}\label{summary}

Putting everything together, we have the following picture. The worlds must split into branch subensembles, their number must be conserved, and within each SG branch, they must separate from each other along the direction of SG splitting without evenly redistributing themselves in each branch. Consequently, repeated SG splittings drive an increasing separation of worlds along the splitting direction. Since each SG stage roughly doubles the inter-world separation along this direction, the growth in the inter-world distances along the splitting direction is exponential.

\subsection{\texorpdfstring{$\mathcal{Q}$}{Q}-Nonlocal Interactions}

What about the $\mathcal{Q} $-locality of the fundamental laws? This too appears plausible, since the two branches no longer interfere once they are sufficiently separated. This is readily accounted for if the inter-world forces attenuate with distance, so that only worlds within the same branch exert significant influence on one another. 

This consideration is not unique to the concatenated SG example. More generally, the $\mathcal{N}$-locality of the hydrodynamic equations ensures that the evolution of the fluid on a region $R_1\subset\mathcal{Q}$ is completely independent of its evolution on a distant, disjoint region $R_2$. Indeed, the evolution of the hydrodynamic fields on $R_1$, given some initial specification of $\rho(\textbf{x}0)$ and $\textbf{v}(\textbf{x},0)$ there, is compatible with a wide range of counterfactual configurations of the hydrodynamic fields on $R_2$. In this sense, the evolution of the Madelung fluid on disjoint regions of $\mathcal{Q}$ is \emph{dynamically autonomous}.  

In the DMIW framework, the underlying inter-world dynamics must reproduce the dynamical autonomy of disjoint regions in the high-density regime. Thus, the worlds located in $R_1$ must behave, to a close approximation, the same regardless of the behavior and distribution of worlds in $R_2$. A particularly simple way to secure this is for the inter-world interactions to decay with distance, so that the influence that disjoint regions exert on each other is negligible. Hence, the $\mathcal{N}$-locality built into the Madelung hydrodynamics strongly suggests that any DMIW interaction reproducing it should be, at the least, $\mathcal{Q}$-local.

A parallel argument applies to decoherence. In realistic systems, environmental interactions continually drives the quantum state into a superposition of approximately orthogonal branches, each supported on disjoint regions of $\mathcal{Q}$, which evolve autonomously because the 
Schr\"odinger dynamics is $\mathcal{N}$-local. Any DMIW model must therefore reproduce the dynamical autonomy of decohered branches. This is naturally achieved if the inter-world forces are $\mathcal{Q}$-local with a range shorter than the typical decoherence-induced separation scale.

While one cannot strictly rule out $\mathcal{Q}$-nonlocal DMIW models, they nonetheless appear poorly motivated, since such models must reproduce the apparent dynamical autonomy of decohered branches, despite treating worlds in distant regions of configuration space as strongly coupled. This would require that the nonlocal interactions somehow cancel on net whenever worlds reside in different branches, and that this cancellation be robust across the enormous variety of branching structures produced by realistic decoherence processes. It seems implausible that a $\mathcal{Q}$-nonlocal DMIW model could reproduce this behavior without introducing additional branch-sensitive structure as a primitive part of the theory or positing extreme fine-tuning of the initial conditions. Both options are unsavory and do not arise naturally from the basic ontology of the DMIW framework.

Taken together with our conclusions from Sect. \ref{summary}, the sparse ontology problem is not just an artifact of the grid-based DMIW model, but a structural feature of any plausible DMIW theory intended to underlie the hydrodynamic picture. Under repeated SG splittings, the worlds in each branch must separate exponentially along the splitting direction. Once they are sufficiently separated, the worlds cease to influence each other due to  the $\mathcal{Q}$-locality of the inter-world forces, and the world ensemble ceases to behave as a coherent whole, leading to deviations from hydrodynamic behavior.

\subsection{Beyond Worlds: Flashes, Fields, and Other Beables}\label{subsec:components}

The preceding analysis has, for clarity, been formulated in terms of DMIW-style worlds, tracing continuous trajectories in configuration space $\mathcal{Q}$. However, the structure of the argument readily extends from the DMIW framework to the broader class of DHV theories, in which the fluid components may differ from point-like worlds that evolve continuously over time.

To illustrate this, consider a DHV in which the fluid is composed not of worlds but of \emph{flashes}, in the spirit of Bell's flash ontology for GRW-type collapse theories.\footnote{As originally conceived in \cite[Ch.~22]{bellSpeakableUnspeakableQuantum2004}, flashes are spacetime events associated with stochastic collapses of the wavefunction. In this context, however, we are not invoking collapse dynamics since the wavefunction is not a fundamental element of DHV theories. Rather, we imagine a potential DHV theory in which the flashes interact directly with each other to reproduce the Madelung evolution.} Flashes are instantaneous events in $\mathcal{Q}$ (or spacetime), rather than persisting world-lines. The hydrodynamic description now interprets $\rho(\mathbf{x},t)$ as the coarse-grained flash density, which obeys a conservation law of the same form as \eqref{continuity3}, for which probabilities are recovered by counting flashes in regions of $\mathcal{Q}$. In this case, decoherence branching is realized as flashes being produced in roughly non-overlapping regions of $\mathcal{Q}$. Successive branching increases the number of independent branch histories without increasing the total number of flashes. Consequently, the expected number of flashes associated with any given branch history decreases, leading to the same effective dilution phenomenon observed in world-based models. This occurs despite the fact that flashes do not persist or split.

One might hope that the problem of sparse ontology can now be avoided by giving up DMIW-style dynamics, since flashes do not follow continuous trajectories and do not satisfy a modified Newtonian law such as Eq.~\eqref{WisemanTrajectory}. However, this does not resolve the issue. Whatever micro-rule is introduced to determine inter-flash influences, it must still account for the dynamical autonomy of decohered branches, which strongly suggests that the interaction rule is $\mathcal{Q}$-local. If the total number of flashes is conserved, repeated branching causes the flash occurrences within each branch to dilute. As the intra-branch flash density falls, the average separation between flashes increases, and---under $\mathcal{Q}$-local interactions---even flashes within the same branch cease to interact strongly enough to sustain the quantum hydrodynamic behavior. Consequently, once the flash density drops below a threshold, the DHV dynamics deviate from standard quantum evolution.

A similar conclusion arises in the quantum field theory (QFT) setting. In one approach to QFT, the state of the system is given by a wave functional, $\psi[\phi]$, over classical field configurations $\phi:\mathbb{R}^3\to\mathbb{R}$.\footnote{An account of the wavefunctional approach is given in \cite{hatfieldQuantumFieldTheory1992,bohmUndividedUniverseOntological1993,hollandQuantumTheoryMotion1993}. Alternatively, one can adopt the particle approach to QFT, in which the quantum state is represented as $\mathcal{Q} = \bigsqcup_{n=0}^\infty Q_n$, i.e.\ the disjoint union of the
$n$-particle configuration spaces $Q_n$ \cite{sebensFundamentalityFields2022,durrBelltypeQuantumField2005}. In that case,  the fluid components correspond to worlds that jump between points belonging to different sectors of
$\mathcal{Q}$ during creation and annihilation events.} A DHV model for the system could be built by replacing $\psi[\phi]$ with an ensemble $\{\phi_k\}$ of discrete interacting fields, that in the high-density regime, behave as a Madelung-like fluid. One could then ground probabilities in direct counting over the $\phi_k$. In this framework, environmental decoherence corresponds to $\psi[\phi]$ splitting into approximately non-overlapping regions in field-configuration space $\mathcal{Q}_\phi$. To ensure the dynamical autonomy of each branch, the inter-field interactions must be $\mathcal{Q}_\phi$-local, so that fields in different regions interact only weakly. Under repeated measurement-induced decoherence, the ensemble $\{\phi_k\}$ thins, and the dynamics tend toward classical-field behavior. This is precisely the same sparse-ontology effect identified in Sect. \ref{sparseworldargument}. 

It must be emphasized that the wavefunctional approach is still under active development—--particularly for fermionic fields---and, to our knowledge, no fully worked-out DHV theory employing field configurations has been formulated.\footnote{See \cite{struyvePilotwaveApproachesQuantum2011} for an assessment of the fermionic wavefunctional approach.} Likewise, a flashed-based DHV in which the dynamics arise solely from inter-flash interactions is highly speculative. Nonetheless, we introduce these sketches to illustrate the point that any complete DHV theory, regardless of the nature of its fluid components, would likely inherit the same abstract structural features that generate the sparse ontology problem in the worlds-based case. The problem is therefore not a peculiarity of DMIW-style worlds, but rather, a generic consequence of the DHV approach.

\section{Sparse Ontology: Feature or Bug?}\label{testability}

Dynamical deviations from standard quantum theory in the low-density regime might seem a virtue rather than a vice, insofar as they render the framework testable. However, one may argue that sparse conditions should already appear in nature, given the enormous number of decoherence events that have occurred throughout the history of the universe. In Sect. \ref{sparseworldargument}, world sparseness is generated artificially by inducing repeated branch splitting via a sequence of SG analyzers. Naturally occurring decoherence branching produces a similar effect since, in the DHV, the ensemble of fluid components partitions into sub-ensembles, each containing fewer components than before the split.\footnote{In contrast with artificially induced splitting, which expands the branch primarily along a single configuration-space direction, decoherence generally produces splittings associated with many distinct sets of degrees of freedom in $\mathcal{Q}$. The high dimensionality of $\mathcal{Q}$ potentially exacerbates the problem of sparse ontology, since small displacements in many different directions may yield a large net effect. To illustrate this, consider two nearby worlds in a given branch, and suppose, in an idealized
scenario, that the wavefunction is approximately separable with respect to each of the $3N$ coordinates of $\mathcal{Q}$.  Assume further that, over the history of the branch, there is at least one decoherence event associated with each coordinate direction, and that such an event stretches the world distribution so that the separation between the pair of worlds
increases by (approximately) an amount $\delta$ along that direction. Then the total increase in separation between the worlds is $\Delta{q}\approx\delta\sqrt{3N}$, which can be enormous if $N$ is the number of particles in the universe. This serves only as a crude heuristic demonstration of the potential influence of a large number of varied decoherence events on the density of worlds.} Furthermore, decoherence ensures that branches remain isolated from each other, thus preventing sub-ensembles from recombining to replenish sparse regions. Since the number of natural branching events plausibly far exceeds what is achievable in the lab, the DHV seems to predict that non-quantum-mechanical behavior should already have been encountered in ordinary circumstances.\footnote{We observe macroscopic objects behaving ``classically'' all around us, but this is not the non-quantum behavior we are referring to. Here, we refer to the microscopic particles that make up macroscopic objects; those particles should behave quantum mechanically according to the standard theory. However, in the DHV, this behavior should be lost when the fluid components become sparse.} 

To explain the apparent quantum character of the world---or a potential null result of an experiment conducted in the lab---the fluid density must be high enough that the number of (artificially or naturally) induced branching events is insufficient to realize the problematic low-density regime. Indeed, the fluid density must be enormous since repeated measurements exponentially increase the average distance between components. As a rough estimate, if we assume that an initial wave packet has a width of $w=.1$mm and the initial inter-world spacing along \(z\) is of order the Planck length, \(l_P\approx1.6\times10^{-35}\,\mathrm{m}\), then approximately 103 SG$_z$ splittings would be required to generate an inter-world separation along $z$ greater than $.1$mm, which would render the packet heavily under-resolved. Thus, very high densities can be probed with a potentially feasible number of repetitions. Even if, however, one posits an initial density that is enormously larger even than Planck-length-separated worlds, to the extent that no realistic sequence of artificial or natural decoherence events could access the low-density regime, then the DHV seems to lose its appeal, since its beyond-quantum inter-component interactions would become fundamentally untestable. \emph{Any} deviation from observed quantum behavior, in other words, eventually implies a \emph{catastrophically large} deviation.

\section{Conclusion}\label{sec13}

The DHV aims to recover quantum theory by replacing the wavefunction with an ensemble of localized fluid components whose collective dynamics underlie the Madelung hydrodynamic evolution in $\mathcal{Q}$. We have argued that the DHV is generically confronted by the problem of sparse ontology, since a plausible DHV will aim to (i) ground probabilities in direct counting over discrete components, and (ii) employ $\mathcal{Q}$–local interactions to reflect the empirical autonomy of decohered branches. These two commitments imply that repeated branching (naturally or artificially induced) thins the component density and drives their dynamics away from standard quantum behavior.

Avoiding the problem of sparse ontology would require abandoning at least one of the DHV's core principles. The issue is that both principles seem essential to the framework. Giving up direct component counting removes the framework's solution to the probability problem, while jettisoning $\mathcal{Q}$-local inter-component interactions makes it unclear how $\mathcal{N}$-local hydrodynamics could be reproduced in the high-density limit. One could instead supplement the discrete ontology with the wavefunction, or an equivalent continuous structure, but this would only serve as a solution to the problem of sparse ontology if the supplementary continuous ontology took over as the primary dynamical element, rendering the inter-component interactions irrelevant. This last point is revealing. Other major approaches to quantum foundations, e.g., Everettian many-worlds, GRW-type collapse models, and Bohmian mechanics, do not encounter the sparsity issue precisely because they keep a continuous wavefunction as the central dynamical entity, allowing quantum structure to be maintained even under repeated decoherence branching. Similarly, CMIW also avoids the problem, since its components are continuously distributed in $\mathcal{Q}$. 

Taken together, these observations suggest that \emph{$\mathcal{Q}$ continuity is not merely a representational convenience, but appears to be a structural requirement of quantum theory}. In this light, the problem of sparse ontology thus provides an argument against discrete hydrodynamic completions of quantum mechanics, and in favor of an ontology that remains continuous at its fundamental level.

\section*{Acknowledgements}

The authors wish to acknowledge support from the University of Vermont (UVM) of various kinds, in the form of start-up funding (from the College of Arts \& Sciences), {\it Mathematica} licensing, and access to the Vermont Advanced Computing Center (VACC).

\begin{appendices}

\section{Discretized CMIW Forces in One and Three Dimensions}
\label{app:CMIW-discretization}

In this appendix, we explicitly show how the one- and three- dimensional DMIW force laws (Eqs.~(\ref{DiscreteTrajectory}) and (\ref{spinforce}) of the main text, respectively) can be derived from the CMIW framework via a suitable application of the finite difference discretization scheme. In each case, the derivation proceeds in two steps: (i) reformulating the hydrodynamic equations in terms of the trajectory-labeling coordinates $\mathbf{C}$, and (ii) applying a finite-difference discretization to the resulting $\mathbf{C}$-derivative expressions.

\subsection{From the 1D CMIW Quantum Force to the Discrete 1D DMIW Quantum Force}
\label{app:1D-force-derivation}

\subsubsection{Step 1: Deriving CMIW from the Hydrodynamic View (HV)}

For a single free spinless particle in one spatial dimension, the hydrodynamic equation of motion, Eq. (\ref{Euler3}), reduces to

    \begin{equation}
        m\,\ddot{x}(t) = -\partial_x Q(x,t),
        \label{eq:1D-hydro-eom}
    \end{equation}
    
\noindent where the quantum potential is

    \begin{equation}
        Q(x,t)
        = -\frac{\hbar^2}{2m}
        \frac{\partial_x^2 \sqrt{\rho(x,t)}}{\sqrt{\rho(x,t)}}.
        \label{eq:1D-quantum-potential}
    \end{equation}
    
In the CMIW approach, the Madelung fluid is constituted by a continuum of world-trajectories, one passing through each point in $\mathcal{Q}$, and satisfying Eq. (\ref{eq:1D-hydro-eom}). Each world carries a permanent label $C$, and since world-trajectories never cross, there is a one-to-one correspondence between $C$ labels and configuration-space points $x \in \mathcal{Q}$ at any time $t$. Thus, the labels $C$ provide a convenient parametrization of $\mathcal{Q}$. It is important to note that, since the $C$ labels are carried by their trajectories, the corresponding parametrization of $Q$ changes with time as the trajectories move through $\mathcal{Q}$.

A particularly useful choice for $C$ is the \emph{uniformizing coordinate}, defined at $t=0$ by the initial cumulative distribution function

    \begin{equation}
        C(x_0) := \int_{-\infty}^{x_0} \rho(x_0',0)\,dx_0',
        \label{eq:uniformizing-def-app}
    \end{equation}

\noindent so that $C\in[0,1]$ labels worlds uniformly with respect to the initial Born density. Because the density $\rho(x,t)$ obeys the continuity equation, cumulative probability is conserved along each world-trajectory. Therefore, although $C$ is assigned to world trajectories only at $t=0$, one has for all later times the identity

    \begin{equation}
        F(x(C,t),t)=C,
    \end{equation}

\noindent where $F(x,t)=\int_{-\infty}^{x}\rho(x',t)\,dx'$ is the instantaneous cumulative distribution and $x(C,t)$ is the world trajectory labeled by $C$. Differentiating this relation with respect to $C$ yields

    \begin{equation}\label{rhoC}
        \rho(x,t) = \frac{1}{\frac{\partial{x}}{\partial{C}}}.
    \end{equation}
    
\noindent Rewriting spatial derivatives in terms of $C$-derivatives, one finds that the quantum force can be expressed purely in terms of $x(C,t)$ and its $C$-derivatives

    \begin{equation}\label{QForce}
        F_Q=\frac{\partial{P}}{\partial{C}},\quad\quad P(C,t)
        =
        \frac{\hbar^2}{4m}\,
        \frac{1}{\bigl(\frac{\partial{x}}{\partial{C}}\bigr)^{2}}\;
            \frac{\partial^2}{\partial{C}^2}\left(\frac{1}{\frac{\partial{x}}{\partial{C}}} \right).
    \end{equation}

\noindent The guidance equation becomes 

    \begin{equation}\label{CMIWguidance}
        m\ddot{x}=\frac{\partial}{\partial{C}}\left[\frac{\hbar^2}{4m}\,
        \frac{1}{\bigl(\frac{\partial{x}}{\partial{C}}\bigr)^{2}}\;
        \frac{\partial^2}{\partial{C}^2}\left( \frac{1}{\frac{\partial{x}}{\partial{C}}} \right)\right],
    \end{equation}

\noindent revealing that all traces of the wavefunction have been effectively removed.

\subsubsection{Step 2: Discretization of the $C$-coordinate.}\label{discretization}
To obtain a discrete DMIW model, we now  the $C$-coordinate explicitly. The $C$ interval $[0,1]$ can be divided into $K$ uniform grid points $C_k=\frac{k-1}{K-1}$ with $1\leq{k}\leq{K}$. Each $C_k$ now labels a discrete world trajectory

    \begin{equation}
        x_k(t) := x(C_k,t).
    \end{equation}
    
Derivatives with respect to $C$ are approximated by finite differences between neighboring worlds. A particularly nice choice is the central finite difference approximation, which, when applied $x$ itself, yields:

    \begin{equation}
        \frac{x(C,t)}{\partial{C}}\bigg |_{C=C_k} \;\approx\; \frac{x_{k+1/2}-x_{k-1/2}}{\,\Delta C}\,\,\footnote{Here we have introduced virtual points at half-integer steps to recover a quantum force that is expressed in terms of points on the integer grid.}
    \end{equation}

\noindent Building up the sequence of derivatives as they appear
in Eq. (\ref{CMIWguidance}), we obtain:

    \begin{equation}
        \begin{aligned}
            \frac{\partial}{\partial{C}}\left( \frac{1}{\frac{\partial{x}_{k}}{\partial{C}}} \right)
            &=
            \frac{1}{x_{k+1} - x_k}
            \;-\;
            \frac{1}{x_k - x_{k-1}},
            \\
            \frac{\partial^2}{\partial{C}^2}\left( \frac{1}{\frac{\partial{x}_{k}}{\partial{C}}} \right)
            &=
            \frac{1}{\Delta C}
            \left[
            \frac{1}{x_{k+3/2} - x_{k+1/2}}
            \;-\;
            \,\frac{2}{x_{k+1/2} - x_{k-1/2}}
            \;+\;
            \frac{1}{x_{k-1/2} - x_{k-3/2}}
            \right]
        \end{aligned}
    \end{equation}

\noindent Substituting these finite-difference expressions into the continuum force \eqref{QForce} and evaluating the  final $C$-derivative by central finite difference, we recover Eq. (\ref{DiscreteTrajectory}) for the discrete quantum force acting on the $k$th trajectory:

    \begin{equation}
        \begin{aligned}
            F_{Q_k}
            =\frac{\hbar^2}{4m}\Bigg\{
            &\frac{1}{(x_{k+1}-x_k)^2}
            \left(
            \frac{1}{x_{k+2}-x_{k+1}}
            -\frac{2}{x_{k+1}-x_k}
            +\frac{1}{x_k-x_{k-1}}
            \right)
            \\[0.5em]
            &-\frac{1}{(x_k-x_{k-1})^2}
            \left(
            \frac{1}{x_{k+1}-x_k}
            -\frac{2}{x_k-x_{k-1}}
            +\frac{1}{x_{k-1}-x_{k-2}}
            \right)
            \Bigg\},
        \end{aligned}
            \label{eq:discrete-1D-quantum-force}
    \end{equation}

\noindent which was alternatively derived in \cite{hallQuantumPhenomenaModeled2014}. 

This completes the derivation of the one-dimensional discrete DMIW force law from the CMIW quantum force expression in 1D. 

\subsection{Multi-Dimensional CMIW and Separable Reduction in Three Dimensions}
\label{app:3D-separable}

We now sketch the analogous continuum-to-discrete logic in three dimensions for a single particle, emphasizing
the separable case relevant to the Stern-Gerlach discussion in Sects. \ref{sec4} and \ref{sparseworldargument}.

\subsubsection{CMIW in Trajectory-Label Coordinates.}

In $n$ dimensions the CMIW trajectories, $\mathbf{x}=\mathbf{x}(\mathbf{C},t)$, are labeled with coordinates $\mathbf{C} = (C_1,\ldots,C_n) \in U \subset \mathbb{R}^n$. If the $\textbf{C}$ coordinates are well chosen so that the map to the standard Cartesian coordinates $\textbf{x}$ is invertible, then a multidimensional version of Eq. (\ref{rhoC}) naturally arises as 

    \begin{equation}
        \rho(\mathbf{x},t)\,\det J(\mathbf{C},t)=1
        \qquad\Longleftrightarrow\qquad
        \rho(\mathbf{x},t)=\frac{1}{\det J}=\det K.
    \end{equation}

\noindent where $J^{i}{}_{j} := \frac{\partial x^{i}}{\partial C_{j}}$ and $K^{j}{}_{i} := (J^{-1})^{j}{}_{i} = \frac{\partial C_{j}}{\partial x^{i}}$ are the corresponding Jacobian and inverse Jacobian matrices.

In these variables, the CMIW quantum force contribution to the trajectory dynamics can be written in terms of $K$ and its $C$-derivatives. In \cite{schiffCommunicationQuantumMechanics2012}, a convenient compact form  was found to be

    \begin{equation}
        m\,\ddot x^{i}(\mathbf{C},t)
        =
        -\frac{\hbar^2}{4m}\,
        \frac{\partial}{\partial C_{m}}
        \!\left(
        K^{m}{}_{j}\,K^{k}{}_{i}\,
        \frac{\partial^{2}K^{\ell}{}_{j}}{\partial C_{k}\,\partial C_{\ell}}
        \right),
    \label{eq:multiD-general-force}
\end{equation}

\noindent which is the multidimensional extension of Eq. (\ref{CMIWguidance}).

\subsubsection{ Separable Initial States and Grid Placement in $\mathbf{C}$.}

From here on, we assume that the initial wavefunction is separable,
$\psi(\mathbf{x},0)=\psi_x(x,0)\psi_y(y,0)\psi_z(z,0)$, so that
$\rho(\mathbf{x},0)=\rho_x(x,0)\rho_y(y,0)\rho_z(z,0)$, and we restrict our attention to systems that have separable Schr\"odinger dynamics. One can then introduce the three uniformizing coordinates (CDF coordinates)

    \begin{equation}
        \begin{aligned}
            C_x(x,t) &= \int_{-\infty}^{x} |\psi_x(x',t)|^2\,dx', \\
            C_y(y,t) &= \int_{-\infty}^{y} |\psi_y(y',t)|^2\,dy', \\
            C_z(z,t) &= \int_{-\infty}^{z} |\psi_z(z',t)|^2\,dz'.
        \end{aligned}
        \label{eq:CDF-coordinates}
    \end{equation}
    
\noindent These imply that the mapping $\mathbf{x}(\mathbf{C},t)$ is coordinate-wise:

    \begin{equation}
        x=x(C_x,t),\qquad y=y(C_y,t),\qquad z=z(C_z,t),
        \label{eq:separable-mapping}
    \end{equation}
    
\noindent so that $J$ and $K$ are diagonal:

    \begin{equation}
        J=\mathrm{diag}\!\left(\frac{\partial x}{\partial C_x},\frac{\partial y}{\partial C_y},\frac{\partial z}{\partial C_z}\right),
        \qquad
        K=\mathrm{diag}\!\left(\frac{\partial C_x}{\partial x},\frac{\partial C_y}{\partial y},\frac{\partial C_z}{\partial z}\right).
        \label{eq:K-diagonal}
    \end{equation}
    
Since $K$ is diagonal, and each diagonal element depends only on its corresponding $C$-coordinate,
Eq.~\eqref{eq:multiD-general-force} separates into three independent equations. The $x$-component becomes

    \begin{equation}
        m\,\ddot x(C_x,t)
        =
        -\frac{\hbar^2}{4m}\,
        \frac{\partial}{\partial C_x}
        \left[
        \left(\frac{\partial C_x}{\partial x}\right)^2
        \frac{\partial^2}{\partial C_x^2}\!\left(\frac{\partial C_x}{\partial x}\right)
        \right],
        \label{eq:55-like-x}
    \end{equation}
    
\noindent with analogous expressions for $y(C_y,t)$ and $z(C_z,t)$. This is the same as the one-dimensional force expression Eq. (\ref{CMIWguidance}).

\subsubsection{Discretization on a 3D $\mathbf{C}$-Lattice.}

Finally, one can discretize $(C_x,C_y,C_z)$ on a uniform grid
$C_{x_i},C_{y_j},C_{z_k}$ and define world positions $x_{i}(t)=x(C_{x_i},t)$, $y_{j}(t)=y(C_{y_j},t)$, $z_{k}(t)=z(C_{z_k},t)$. By construction, the grid on $\textbf{C}$-space induces a grid on $\mathcal{Q}$ with fixed values of $i,j,$ or $k$ picking out \emph{levels} of worlds at constant $x,y$ and $z$ values, respectively, in the Cartesian grid. Furthermore, because the separable reduction yields three one-dimensional equations, each component can be discretized using the same one-dimensional finite-difference force stencil as in App. \ref{discretization}, applied independently along $i$, $j$, and $k$. Each separate equation governs the evolution of the discretized $i$-, $j$-, and $k$-levels, respectively.

\section{Fixing the ``Flux Diminution Problem'' in SG-Interferometer Measurements}\label{secB}

\subsection{Introduction}

Sect.~\ref{sparseworldargument} presents a ``proof of concept'' thought experiment---using 
ordinary, well-understood experimental equipment---whose outcomes, were it to be performed, 
are not in much doubt. Nevertheless, the proposed experiment (or something like it) is not 
actually realizable in practice. The reason is the ``flux diminution problem'': for each
SG measurement that is applied, there is a 50\% chance that the ``down'' outcome is 
achieved---resulting in a spot on the absorber screen, \emph{and the end of the experiment.}
Thus, at each such measurement, the experiment runs a substantial risk of immediate cessation. Since,
according to  Sect.~\ref{testability}, the particle must undergo something like $N=100$ such successive SG measurements in order to achieve extreme under-resolution of the wave packet and completely rule out even Planck-length-scale world separations, the likelihood of ever making it this far is only
one in $2^{100}$, or about $8\times10^{-31}$. Thus, the very rarefaction that leads to the extreme 
world spreading needed to verify the sparse-world effect, also ensures that such an outcome
is never actually observed in practice.

A more compelling argument can be crafted, however, using a variant of the above experiment 
that suffers from no such flux diminution problem, and might therefore be realizable in practice. 
The mechanics are a bit more complex than what is presented in Sect. 5, however, which is why 
we did not present this version in the main body of the paper---i.e., because from a purely 
conceptual standpoint, it is not necessary. Even here, in the appendix, we do not delve into
the tedium of the many experimental details that would likely have to be sorted out to make
it all actually work using current technology. Rather, we simply wish
to make a plausibility argument as to why and how such an experiment could indeed be realized
in practice.

To avoid the reduction in particle flux after each stage, the up-and-down-branch wave packets that emerge must be realigned in \emph{physical} space (e.g., by using SG analyzers ``in reverse''). If this were all there were to it, however, it would not suffice; for this would cause the world sub-ensembles to merge, thereby preventing their thinning in {$\mathcal{Q}$ space}.\footnote{For convenience, we carry out the analysis in terms of worlds, but the argument applies equally well to the broader class of DHV components.} In effect, the realignment would simply ``undo'' the original SG operation, returning us to our starting point. To circumvent this difficulty, each SG stage can be followed by a non-projective measurement inserted after the SG split. The measurement process entangles the system with the degrees of freedom of the detection device, so that even though the branches are subsequently recombined in \emph{physical} space, they remain well separated in $\mathcal{Q}$ space. 

The above arrangement ensures that the world ensembles continue to thin as they progress through the SG sequence. \emph{Moreover, there is no fear of the experiment ``ending'' after each SG measurement is performed}, since there are no absorbers/screens. \emph{Regardless} of the particular outcome of any given measurement, i.e. ``spin up'' or ``spin down'', the experiment continues. Over time, all $2^N$ possible historical outcomes are duly recorded (across all worlds/branches), in the set of $N$ distinct non-destructive detectors. Note that for
this scheme to work, it is essential that each detector outcome is preserved over time thereafter. Thus, for example, it is not possible (at least not without further refinements) to work with just a single detector that gets ``reset'' between each 
successive SG measurement. Indeed, it is clear, by the same token, that any fewer than $N$ detectors will also fail. But a single experimental run with $N\approx 100$ detectors is feasible---in contrast to running $2^{100}$ experiments with a single detector, as per Sect.~\ref{sparseworldargument}, which is \emph{not} a feasible proposal. Moreover, in the single-run scenario proposed here, all that is \emph{really} 
needed is $\approx 100$ distinct recorded measurement \emph{outcomes}---all of which, in principle, could
be obtained from the same detector. This suggests that indeed, the proposed scheme
should be realizable.

\subsection{Modeling the Detector}
\label{detectormodel}

We consider a minimally invasive measurement that detects which branch the particle is in after it has passed through an SG analyzer. To this end, we introduce window functions $w_{\uparrow}$ and $w_{\downarrow}$ defined as

   \begin{equation}
        w_{\uparrow}(\vec{x}) =
        \begin{cases}
        1, & \text{if } \vec{x} \in R_{\uparrow}, \\[6pt]
        0, & \text{otherwise,}
        \end{cases}
        \qquad
        w_{\downarrow}(\vec{x}) =
        \begin{cases}
        1, & \text{if } \vec{x} \in R_{\downarrow}, \\[6pt]
        0, & \text{otherwise,}
        \end{cases}
\end{equation}

\noindent supported in disjoint regions, $R_{\uparrow}$ and $R_{\downarrow}$, where the up and down branches are expected to pass.
The detector is idealized as a single particle of mass $M\gg m$, with coordinates $\vec{x}_d = (x_d, y_d, z_d)$. The detection process is modeled by the interaction Hamiltonian

    \begin{equation}
        \hat{H}_{\text{int}}=\lambda(t)\begin{pmatrix}
        a\,w_{\uparrow}(\vec{x}) & 0\\[4pt]
        0 & -a\,w_{\downarrow}(\vec{x}),
        \end{pmatrix}\hat{p}_{z_d},
    \end{equation}

\noindent where $a$ is a constant that determines the magnitude of the detector’s translation.
The function $\lambda(t)$ is a real-valued switching function that modulates the strength of the particle–detector interaction, specifying the temporal window during which the measurement device is active.
For simplicity, we idealize it as a square pulse,

\begin{equation}
    \lambda(t)=
    \begin{cases}
    \lambda_0, & t_i \le t \le t_f,\\[4pt]
    0, & \text{otherwise},
    \end{cases}
\end{equation}
corresponding to an interaction that is turned on at $t_i$ and switched off at $t_f$. 

If the detector mass $M$ is sufficiently large that its Laplacian evolution can be neglected, then the initial particle–detector state

    \begin{equation}\label{initial2}
        \Psi(\vec{x}, \vec{x}_d, t_i)
        = \frac{1}{\sqrt{2}}
        \begin{pmatrix}
        \phi_{\uparrow}(\vec{x},t_i) \\[4pt]
        \phi_{\downarrow}(\vec{x},t_i)
        \end{pmatrix}
        D_0(\vec{x}_d), 
    \end{equation}  

\noindent (where $ D_0(\vec{x}_d)$ denotes the detector state initially localized around $z_{d_0}$), evolves according to 

    \begin{equation}
            \nonumber\Psi(\vec{x}, \vec{x}_d, t_f)=\frac{1}{\sqrt{2}}\exp\left[-\frac{i\Delta{t}}{\hbar}\left(-\frac{\hbar^2}{2m}\nabla^2+\hat{H}_{\text{int}}\right)\right]\begin{pmatrix}
        \phi_{\uparrow}(\vec{x},t_i)  \\[4pt]
        \phi_{\downarrow}(\vec{x},t_i) 
        \end{pmatrix}D_0(\vec{x}_d).
    \end{equation}
    
\noindent If $\phi_{\uparrow/\downarrow}(\vec{x},t)$ lie exclusively within, or exclusively without, its corresponding detector region $R_{\uparrow/\downarrow}$ while the detector is active, then the exponential operator approximately factorizes\footnote{If $\phi_{\uparrow/\downarrow}(\vec{x},t)$ are supported entirely within (or outside) of $R_{\uparrow/\downarrow}$, then factorization is exact by the Baker-Campbell-Hausdorff formula, since the commutator $[\nabla^2,w(\vec{x})]$ acts trivially on $\phi_{\uparrow/\downarrow}(\vec{x},t)$. In assuming approximate factorization, we neglect the marginal case in which the high-amplitude regions of $\phi_{\uparrow/\downarrow}(\vec{x},t)$ overlap with the boundaries of $R_{\uparrow/\downarrow}$ while the detector is on.}  yielding

\begin{equation}
        \begin{aligned}
            \nonumber\Psi(\vec{x}, \vec{x}_d, t_f)&\approx\frac{1}{\sqrt{2}}\exp\left[-\frac{i\Delta{t}}{\hbar}\left(-\frac{\hbar^2}{2m}\nabla^2\right)\right]\exp\left[-\frac{i\Delta{t}}{\hbar}\hat{H}_{\text{int}}\right]\begin{pmatrix}
        \phi_{\uparrow}(\vec{x},t_i)  \\[4pt]
        \phi_{\downarrow}(\vec{x},t_i) 
        \end{pmatrix}D_0(\vec{x}_d).
            \\
            &=\frac{1}{\sqrt{2}}\exp\left[-\frac{i\Delta{t}}{\hbar}\hat{H}_{\text{int}}\right]D_0(\vec{x}_d)\exp\left[-\frac{i\Delta{t}}{\hbar}\left(-\frac{\hbar^2}{2m}\nabla^2\right)\right]\begin{pmatrix}
        \phi_{\uparrow}(\vec{x},t_i)  \\[4pt]
        \phi_{\downarrow}(\vec{x},t_i) 
        \end{pmatrix}
            \\
            &=\frac{1}{\sqrt{2}}\exp\begin{pmatrix}
                -\lambda_0{\Delta{t}}a\,w_{\uparrow}(\vec x)\partial_{z_d} & 0\\[4pt]
                 0 & \lambda_0{\Delta{t}}a\,w_{\downarrow}(\vec x)\partial_{z_d}
            \end{pmatrix}D_0(\vec{x}_d)\begin{pmatrix}
        \phi_{\uparrow}(\vec{x},t_f)  \\[4pt]
        \phi_{\downarrow}(\vec{x},t_f) 
        \end{pmatrix}
        \\
            &=\frac{1}{\sqrt{2}}\left[\begin{pmatrix}
                \exp\left[-\lambda_0{\Delta{t}}a\,w_{\uparrow}(\vec x)\partial_{z_d}\right]& 0\\[4pt]
                 0 & \exp\left[\lambda_0{\Delta{t}}a\,w_{\downarrow}(\vec x)\partial_{z_d}\right]
            \end{pmatrix}\right]D_0(\vec{x}_d)\begin{pmatrix}
        \phi_{\uparrow}(\vec{x},t_f)  \\[4pt]
        \phi_{\downarrow}(\vec{x},t_f) 
        \end{pmatrix}
        \\
        &=\frac{1}{\sqrt{2}}\left[\begin{pmatrix}
                \exp\left[-\lambda_0{\Delta{t}}a\,w_{\uparrow}(\vec x)\partial_{z_d}\right] D_0(\vec{x}_d) & 0\\[4pt]
                 0 & \exp\left[\lambda_0{\Delta{t}}a\,w_{\downarrow}(\vec x)\partial_{z_d}\right] D_0(\vec{x}_d)
            \end{pmatrix}\right]\begin{pmatrix}
        \phi_{\uparrow}(\vec{x},t_f) \\[4pt]
        \phi_{\downarrow}(\vec{x},t_f)
        \end{pmatrix}.
        \end{aligned}
    \end{equation}
    
\noindent Using the identity $e^{\pm{c\partial_z}}f(x,y,z)=f(x,y,z\pm{c})$, gives 

    \begin{align}
         \Psi(\vec{x},\vec{x}_d,t_f)
         &=\frac{1}{\sqrt{2}}\,
        \begin{pmatrix}
        \phi_{\uparrow}(\vec{x},t_f)\\[6pt]
        0
        \end{pmatrix}
        D_0\!\big(x_d,y_d,\,z_d-\lambda_0 \Delta{t}\,a\,w_{\uparrow}(\vec x)\big)
        \notag\\[6pt]
        &\quad+\frac{1}{\sqrt{2}}\,
        \begin{pmatrix}
        0\\[6pt]
        \phi_{\downarrow}(\vec{x},t_f)
        \end{pmatrix}
        D_0\!\big(x_d,y_d,\,z_d+\lambda_0 \Delta{t}\,a\,w_{\downarrow}(\vec x)\big)
        \notag\\[6pt]
        &:=\frac{1}{\sqrt{2}}\,
        \begin{pmatrix}
        \phi_{\uparrow}(\vec{x},t_f)\\[6pt]
        0
        \end{pmatrix}
        D_\uparrow\!\big(\vec{x},\vec{x}_d\big)
        +\frac{1}{\sqrt{2}}\,
        \begin{pmatrix}
        0\\[6pt]
        \phi_{\downarrow}(\vec{x},t_f)
        \end{pmatrix}
        D_\downarrow\!\big(\vec{x},\vec{x}_d\big)
        \label{eq:psi-final}
    \end{align}

The high-amplitude domain of $\phi_{\uparrow/\downarrow}(\vec{x},t_f)D_{\uparrow/\downarrow}\!\big(\vec{x},\vec{x}_d\big)$ occurs where the high-amplitude domains of $\phi_{\uparrow/\downarrow}(\vec{x},t_f)$ and $D_{\uparrow/\downarrow}\!\big(\vec{x},\vec{x}_d\big)$ overlap. While $\phi_{\uparrow/\downarrow}(\vec{x},t)$ passes through $R_{\uparrow/\downarrow}$---that is, while the up/down-packet passes through the corresponding up/down-window---$\lvert\phi_{\uparrow/\downarrow}(\vec{x},t_f)D_{\uparrow/\downarrow}\!\big(\vec{x},\vec{x}_d\big)\rvert^2$ is sharply peaked near $z_d=z_{d_0}\pm{\lambda_0}a\Delta{t}$, indicating that the detector registers the particle's passage through $R_{\uparrow/\downarrow}$. Conversely, if $\phi_{\uparrow/\downarrow}(\vec{x},t_f)$ somehow misses its window, then $\lvert\phi_{\uparrow/\downarrow}(\vec{x},t_f)D_{\uparrow/\downarrow}\!\big(\vec{x},\vec{x}_d\big)\rvert^2$ remains sharply peaked around $z_{d_0}$, indicating that the detector has not registered the particle. However, this serves as a ``safeguard'' only; the experimental parameters should have been chosen such that this eventuality does not occur.
\footnote{That said, in Sect.~\ref{reset}, we will make explicit use of this feature to reset the detector to its ready state.}

In the DMIW framework, worlds are represented by configuration trajectories $\textbf{x}_k(t)=(\vec{x}_k(t),\vec{x}_{d_k}(t))$, which encode the behavior of both the particle and the detector. During the measurement interaction, the world ensemble divides into high-density sub-ensembles located where $\lvert\phi_{\uparrow/\downarrow}(\vec{x},t_f)D_{\uparrow/\downarrow}\!\big(\vec{x},\vec{x}_d\big)\rvert^2$ is peaked. In worlds belonging to the up/down sub-ensemble, the detector is displaced by approximately $\pm\lambda_0{a}\Delta{t}$ along $z_d$ when the particle is in $R_{\uparrow/\downarrow}$, respectively. Conversely---and again, as a 
safeguard---the detector remains stationary in worlds where the particle somehow misses either window.

\subsection{Optional Step: Recording all Detector Outcomes}
\label{optional}

We are not experimentalists; however, our experimentalist colleagues have convinced us that particle detectors can be expensive.  In contrast, recording devices, designed to
record the (macroscopic) outcome of a detector, may be comparatively inexpensive. 
That being the case, to save on costs---but at the expense of complicating the
experiment somewhat---we suggest here an additional ``optional'' step, whereby
the detector outcome is macroscopically recorded after each SG splitting, using
a recording device.  The detector state itself may then be safely reset 
to its ready state, in between
each SG measurement---thereby enabling a single detector to be used for all $N \approx 100$
measurements. 

Of course, for this scheme to work, it is essential that a \emph{different} recording device, denoted $E^j$, be used for each SG measurement, $j$. Once 
the outcome of measurement $j$ is recorded, the corresponding recorder $E^j$ is permanently decoupled from the rest of the experimental apparatus, 
in such a way that its state is preserved thereafter over time. 
In this manner, the sequence of recorded measurement outcomes itself,
stored across the $N \approx 100$ recording devices, ensures separation of
the $2^N$ distinct branches. 

In basic terms, the scheme may be implemented as follows. After the detector
has performed its $j$'th SG measurement, the detector is placed in contact with
recorder $E(j)$, with which its state becomes correlated (following a procedure
similar to that of Sect.~\ref{detectormodel}, but involving only macroscopic 
coordinates). Denoting the coordinates for recorder $E^j$ as $\vec{x}_j$, we find that the final wavefunction takes the form

    \begin{equation}
    \label{recorder}
        \Psi(\vec{x},\vec{x}_d,\vec{x}_j)=\frac{1}{\sqrt{2}}\,
        \begin{pmatrix}
        \phi_{\uparrow}(\vec{x},t_f)\\[6pt]
        0
        \end{pmatrix}
        D_\uparrow\!\big(\vec{x},\vec{x}_d\big)E^j_\uparrow(\vec{x}_j)
        +\frac{1}{\sqrt{2}}\,
        \begin{pmatrix}
        0\\[6pt]
        \phi_{\downarrow}(\vec{x},t_f)
        \end{pmatrix}
        D_\downarrow\!\big(\vec{x},\vec{x}_d\big)E^j_\downarrow(\vec{x}_j),
    \end{equation}  
where $E^j_{\uparrow/\downarrow}(\vec{x}_j)$ represents the up/down state
for recorder $E^j$. 

Of course, even Eq.~(\ref{recorder}) represents a 
simplification only, as we have left out all of the $\vec{x}_{{j'}<j}$ coordinates
from the wavefunction, corresponding to recordings for previous measurements.
If these were included, they would manifest explicitly all $2^j$ wavefunction branches that arise after the first $j$ measurements have been
performed. We have
also neglected all of the $\vec{x}_{{j'}>j}$ coordinates, for the as-yet-unused recorders
$E^{{j'}>j}$; however, these merely contribute additional factors of 
$E_0^{j'}(\vec{x}_{j'})$ into the overall wavefunction, where $E_0^{j'}(\vec{x}_{j'})$ represents the ready state of recorder $E^{j'}$.

Note that the entanglement between the detector and the many recorder degrees of freedom is what ensures that the branches remain decoherent after the measurement process concludes. Accordingly, $E^j_{\uparrow}(\vec{x}_j)$ and $E_{\downarrow}(\vec{x}_j)$ act as record states that prevent the branches from reuniting in $\mathcal{Q}$---even if the particle states $\phi_\uparrow(\vec{x},t)$ and $\phi_\downarrow(\vec{x},t)$ (and corresponding detector states) are later 
realigned---which is what we consider next. 

\subsection{Recombining the Particle Wave Packet Branches}
\label{recombining}

 The particle wave packet branches, $\phi_\uparrow(\vec{x},t)$ and $\phi_\downarrow(\vec{x},t)$, are initially accelerated away from each other in opposite directions along the $z$-direction as they traverse the SG analyzer. After exiting the magnetized region, they continue to propagate approximately inertially in opposite $z$-directions,
 until they reach their respective disjoint regions of physical space---i.e. 
 $R_\uparrow$ and $R_\downarrow$, respectively. 
 To realign the packets so that they are once again centered along the $y$-axis, 
 their motion along $z$ must first be reversed. This can be achieved by sending the packets through a second SG analyzer whose field gradient is opposite that of the first SG analyzer.\footnote{Note that when the wave packets are at their most separated, their distance from one another
may be much larger than the inside of one SG analyzer. If so, this may require
\emph{two} new SG analyzers to implement the deceleration, one for each particle 
wave packet branch.}

Since the field configuration of the initial analyzer is taken to be $\vec{B}=(0,0,bz)$, it suffices for the second analyzer to be identical to the first but rotated 180$^{\circ}$ about the $y$-axis such that its $z$-orientation is flipped relative to the first analyzer. The reversed field gradient accelerates the packets in directions opposite to those induced by the first SG analyzer, redirecting them toward the $y$-axis. Once the packets exit the second analyzer and are traveling towards each other, they must be decelerated along the $z$-direction so that their $z$-velocities vanish as they reach the $y$-axis. This can be accomplished by introducing yet another SG analyzer, producing the same field gradient as the first. 

After realignment, the particle wave packets are once again essentially in their initial state\footnote{Of course, the particle state has now also become entangled with the most recent detector and/or recorder states.}---meaning their position and motion are confined to a third disjoint region, $R_0$, lying near $x=0,z=0$.
The region $R_0$ is identical to that occupied by the particle prior to the previous
SG measurements, apart from a translation in the $y$ direction. The particle
is thus prepared to undergo the next stage of SG measurement.
 
\subsection{Optional Step: Resetting the Detector}
\label{reset}

If a different detector is being used for each SG measurement, then it is
presumed that the new detector, to be used for the $(j+1)$'th measurement, 
is initialized in its ready state, $D_0(\vec{x}_d)$. Ignoring the other detector
degrees of freedom, then, after applying Sect. \ref{recombining}, then, 
we have restored the particle-plus-new-detector wavefunction to the initial form
of Eq. (\ref{initial2}), and are thus ready to begin a new SG measurement. 

If, on the other hand, we have adopted the  ``cost-cutting'' implementation 
of Sect. B.3, then our particle-plus-detector wavefunction
is still of the form of Eq. (\ref{eq:psi-final}). Thus, the detector is
still in its $D_{\uparrow/\downarrow}(\vec{x}_d)$ measured outcome states,
as opposed to its ready state, $D_0(\vec{x}_d)$. That said---and unlike 
the case in Sect. \ref{detectormodel}---the particle wavefunctions are
at this point no longer situated in the regions $R_{\uparrow/\downarrow}$, 
but in the disjoint region, $R_0$.  

The above suggests a very straightforward method for resetting the
detector back to its ready state---i.e., to simply perform another
detector measurement by turning on $\hat{H}_{\text{int}}$.  This
time, however, because the particle is \emph{not} to be found 
in the regions $R_{\uparrow/\downarrow}$, the $w_{\uparrow/\downarrow}(\vec x)$
contributions in Eq. (\ref{eq:psi-final}) are zero, and so 
the high-amplitude region of $\lvert\phi_{\uparrow/\downarrow}(\vec{x},t)D_{\uparrow/\downarrow}\!\big(\vec{x},\vec{x}_d\big)\rvert^2$ occurs where $D_{\uparrow/\downarrow}\!\big(\vec{x},\vec{x}_d\big)=D_0(\vec{x}_d)$. This indicates that, after performing this new detector measurement, the detector is (with extremely high likelihood) once again to be found in its ready 
configuration.

\subsection{Summary of the SG Sequence}

In order to convert the procedure as described in Sect. \ref{sparseworldargument}
into something more feasible for an actual laboratory experiment, each stage of the SG sequence must be modified by appending a Welcher Weg detector measurement (or two), followed by two (or three) realigning SG analyzers (and optionally, a sequence of recorder measurements). The repeated Welcher Weg measurements (and ensuing recorder measurements), generate an iterated branching structure within the wavefunction. This structure is, in general, quite intricate, since at every stage the particle's spinor state becomes entangled with new detector (and recorder) degrees of freedom, causing the wavefunction to split into tendrils that spread throughout $\mathcal{Q}$. Despite the complexity of this branching in the environmental coordinates, the recurrent splitting and realigning of the wave packets along $x$ and $z$ drive the branches to oscillate along those directions. In the DMIW framework, this appears as the world ensemble dividing into sub-ensembles that propagate in various directions in $\mathcal{Q}$ but remain systematically oscillatory along $x$ and $z$ at each repetition. Consequently, the average inter-world separation along $x$ and $z$ is successively doubled. Hence, the thinning effect characteristic of the original SG sequence is preserved, even after the flux-conserving modifications are introduced---thus achieving the desired aim.

\end{appendices}

\bibliographystyle{unsrt}
\bibliography{references}

@incollection{albertElementaryQuantumMetaphysics1996,
  title = {Elementary {{Quantum Metaphysics}}},
  booktitle = {Bohmian {{Mechanics}} and {{Quantum Theory}}: {{An Appraisal}}},
  author = {Albert, David Z.},
  editor = {Cushing, James T. and Fine, Arthur and Goldstein, Sheldon},
  year = 1996,
  pages = {277--284},
  publisher = {Springer Netherlands},
  address = {Dordrecht},
  doi = {10.1007/978-94-015-8715-0_19},
  urldate = {2025-10-16},
  isbn = {978-94-015-8715-0},
  langid = {english}
}

@article{alloriCommonStructureBohmian2008,
  title = {On the {{Common Structure}} of {{Bohmian Mechanics}} and the {{Ghirardi}}--{{Rimini}}--{{Weber Theory}}},
  author = {Allori, Valia and Goldstein, Sheldon and Tumulka, Roderich and Zangh{\`i}, Nino},
  year = 2008,
  month = sep,
  journal = {The British Journal for the Philosophy of Science},
  publisher = {The University of Chicago Press},
  doi = {10.1093/bjps/axn012},
  urldate = {2025-10-16},
  copyright = {\copyright{} 2008 by The Authors. All rights reserved.},
  langid = {english}
}

@book{bellSpeakableUnspeakableQuantum2004,
  title = {Speakable and {{Unspeakable}} in {{Quantum Mechanics}}: {{Collected Papers}} on {{Quantum Philosophy}}},
  shorttitle = {Speakable and {{Unspeakable}} in {{Quantum Mechanics}}},
  author = {Bell, J. S.},
  year = 2004,
  edition = {2},
  publisher = {Cambridge University Press},
  address = {Cambridge},
  doi = {10.1017/CBO9780511815676},
  urldate = {2025-10-10},
  isbn = {978-0-521-81862-9}
}

@article{bohmSuggestedInterpretationQuantum1952,
  title = {A {{Suggested Interpretation}} of the {{Quantum Theory}} in {{Terms}} of "{{Hidden}}" {{Variables}}. {{I}}},
  author = {Bohm, David},
  year = 1952,
  journal = {Physical Review},
  volume = {85},
  number = {2},
  pages = {166--179},
  doi = {10.1103/PhysRev.85.166}
}

@book{bohmUndividedUniverseOntological1993,
  title = {The {{Undivided Universe}}: {{An Ontological Interpretation}} of {{Quantum Theory}}},
  shorttitle = {The {{Undivided Universe}}},
  author = {Bohm, David and Hiley, Basil J.},
  year = 1993,
  publisher = {Routledge},
  address = {London},
  urldate = {2025-11-19},
  langid = {english}
}

@article{deutschQuantumTheoryProbability1999,
  title = {Quantum Theory of Probability and Decisions},
  author = {Deutsch, David},
  year = 1999,
  month = aug,
  journal = {Proceedings of the Royal Society of London. Series A: Mathematical, Physical and Engineering Sciences},
  publisher = {The Royal Society},
  doi = {10.1098/rspa.1999.0443},
  urldate = {2025-10-13},
  langid = {english}
}

@article{durrBelltypeQuantumField2005,
  title = {Bell-Type Quantum Field Theories},
  author = {D{\"u}rr, Detlef and Goldstein, Sheldon and Tumulka, Roderich and Zangh{\`i}, Nino},
  year = 2005,
  month = jan,
  journal = {Journal of Physics A: Mathematical and General},
  volume = {38},
  number = {4},
  pages = {R1},
  issn = {0305-4470},
  doi = {10.1088/0305-4470/38/4/R01},
  urldate = {2025-11-21},
  langid = {english}
}

@misc{durrBohmianMechanicsMeaning1995,
  title = {Bohmian {{Mechanics}} and the {{Meaning}} of the {{Wave Function}}},
  author = {D{\"u}rr, Detlef and Goldstein, Sheldon and Zangh{\`i}, Nino},
  year = 1995,
  month = dec,
  number = {arXiv:quant-ph/9512031},
  eprint = {quant-ph/9512031},
  publisher = {arXiv},
  doi = {10.48550/arXiv.quant-ph/9512031},
  urldate = {2026-01-30},
  archiveprefix = {arXiv}
}

@article{durrQuantumEquilibriumOrigin1992,
  title = {Quantum Equilibrium and the Origin of Absolute Uncertainty},
  author = {D{\"u}rr, Detlef and Goldstein, Sheldon and Zangh{\'i}, Nino},
  year = 1992,
  month = jun,
  journal = {Journal of Statistical Physics},
  volume = {67},
  number = {5},
  pages = {843--907},
  issn = {1572-9613},
  doi = {10.1007/BF01049004},
  urldate = {2025-10-08},
  langid = {english}
}

@article{fabbriDiracTheoryHydrodynamic2023,
  title = {Dirac {{Theory}} in {{Hydrodynamic Form}}},
  author = {Fabbri, Luca},
  year = 2023,
  month = may,
  journal = {Foundations of Physics},
  volume = {53},
  number = {3},
  pages = {54},
  issn = {1572-9516},
  doi = {10.1007/s10701-023-00695-w},
  urldate = {2026-01-18},
  langid = {english}
}

@article{hallQuantumPhenomenaModeled2014,
  title = {Quantum {{Phenomena Modeled}} by {{Interactions}} between {{Many Classical Worlds}}},
  author = {Hall, Michael J. W.},
  year = 2014,
  journal = {Physical Review X},
  volume = {4},
  number = {4},
  doi = {10.1103/PhysRevX.4.041013}
}

@book{hatfieldQuantumFieldTheory1992,
  title = {Quantum {{Field Theory Of Point Particles And Strings}}},
  author = {Hatfield, Brian},
  year = 1992,
  series = {Frontiers of {{Physics}}},
  volume = {75},
  publisher = {Addison-Wesley},
  address = {Redwood City, CA},
  urldate = {2025-11-19},
  langid = {english}
}

@misc{herrmannEigenstatesManyInteracting2023,
  title = {Eigenstates in the {{Many Interacting Worlds}} Approach: {{Ground}} States in {{1D}} and {{2D}} and Excited States in {{1D}} (Long Version)},
  shorttitle = {Eigenstates in the {{Many Interacting Worlds}} Approach},
  author = {Herrmann, Hannes and Hall, Michael J. W. and Wiseman, Howard M. and Deckert, Dirk-Andr{\'e}},
  year = 2023,
  month = jul,
  number = {arXiv:1712.01918},
  eprint = {1712.01918},
  primaryclass = {quant-ph},
  publisher = {arXiv},
  doi = {10.48550/arXiv.1712.01918},
  urldate = {2026-01-16},
  archiveprefix = {arXiv}
}

@book{hollandQuantumTheoryMotion1993,
  title = {The {{Quantum Theory}} of {{Motion}}: {{An Account}} of the de {{Broglie-Bohm Causal Interpretation}} of {{Quantum Mechanics}}},
  shorttitle = {The {{Quantum Theory}} of {{Motion}}},
  author = {Holland, Peter R.},
  year = 1993,
  publisher = {Cambridge University Press},
  address = {Cambridge},
  doi = {10.1017/CBO9780511622687},
  urldate = {2025-10-15},
  isbn = {978-0-521-48543-2}
}

@article{hollandUniquenessConservedCurrents2003,
  title = {{Uniqueness of conserved currents in quantum mechanics}},
  author = {Holland, P.},
  year = 2003,
  journal = {Annalen der Physik},
  volume = {515},
  number = {7-8},
  pages = {446--462},
  issn = {1521-3889},
  doi = {10.1002/andp.200351507-805},
  urldate = {2026-01-23},
  copyright = {Copyright \copyright{} 2003 WILEY-VCH Verlag GmbH \& Co. KGaA, Weinheim},
  langid = {ngerman}
}

@article{hsuSternGerlachDynamicsQuantum2011,
  title = {Stern-{{Gerlach}} Dynamics with Quantum Propagators},
  author = {Hsu, Bailey C.},
  year = 2011,
  journal = {Physical Review A},
  volume = {83},
  number = {1},
  doi = {10.1103/PhysRevA.83.012109}
}

@article{krekelsZigzagDynamicsStern2024,
  title = {Zig-Zag Dynamics in a {{Stern}}--{{Gerlach}} Spin Measurement},
  author = {Krekels, Simon and Maes, Christian and Meerts, Kasper and Struyve, Ward},
  year = 2024,
  month = mar,
  journal = {Proceedings of the Royal Society A: Mathematical, Physical and Engineering Sciences},
  volume = {480},
  number = {2285},
  pages = {20230861},
  publisher = {Royal Society},
  doi = {10.1098/rspa.2023.0861},
  urldate = {2025-10-27}
}

@article{lombardiniInteractingQuantumTrajectories2024,
  title = {Interacting Quantum Trajectories for Particles with Spin 1/2},
  author = {Lombardini, R. and Poirier, B.},
  year = 2024,
  month = aug,
  journal = {Molecular Physics},
  publisher = {Taylor \& Francis},
  issn = {0026-8976},
  urldate = {2025-09-30},
  copyright = {\copyright{} 2024 Informa UK Limited, trading as Taylor \& Francis Group},
  langid = {english}
}

@article{madelungQuantentheorieHydrodynamischerForm1927,
  title = {{Quantentheorie in hydrodynamischer Form}},
  author = {Madelung, E.},
  year = 1927,
  month = mar,
  journal = {Zeitschrift f\"ur Physik},
  volume = {40},
  number = {3},
  pages = {322--326},
  issn = {0044-3328},
  doi = {10.1007/BF01400372},
  urldate = {2025-07-31},
  langid = {ngerman}
}

@book{neyWaveFunctionEssays2013,
  title = {The {{Wave Function}}: {{Essays}} on the {{Metaphysics}} of {{Quantum Mechanics}}},
  shorttitle = {The {{Wave Function}}},
  editor = {Ney, Edited by Alyssa and Albert, David Z.},
  year = 2013,
  month = apr,
  publisher = {Oxford University Press},
  address = {Oxford, New York},
  isbn = {978-0-19-979080-7}
}

@article{plattModernAnalysisStern1992,
  title = {A Modern Analysis of the {{Stern}}--{{Gerlach}} Experiment},
  author = {Platt, Daniel E.},
  year = 1992,
  month = apr,
  journal = {American Journal of Physics},
  volume = {60},
  number = {4},
  pages = {306--308},
  issn = {0002-9505},
  doi = {10.1119/1.17136},
  urldate = {2025-10-14}
}

@article{poirierBohmianMechanicsPilot2010,
  title = {Bohmian Mechanics without Pilot Waves},
  author = {Poirier, Bill},
  year = 2010,
  month = may,
  journal = {Chemical Physics},
  series = {Dynamics of Molecular Systems: {{From}} Quantum to Classical},
  volume = {370},
  number = {1},
  pages = {4--14},
  issn = {0301-0104},
  doi = {10.1016/j.chemphys.2009.12.024},
  urldate = {2025-10-23}
}

@article{reddigerMathematicalTheoryMadelung2023,
  title = {Towards a Mathematical Theory of the {{Madelung}} Equations: {{Takabayasi}}'s Quantization Condition, Quantum Quasi-Irrotationality, Weak Formulations, and the {{Wallstrom}} Phenomenon},
  shorttitle = {Towards a Mathematical Theory of the {{Madelung}} Equations},
  author = {Reddiger, Maik and Poirier, Bill},
  year = 2023,
  month = apr,
  journal = {Journal of Physics A: Mathematical and Theoretical},
  volume = {56},
  number = {19},
  pages = {193001},
  publisher = {IOP Publishing},
  issn = {1751-8121},
  doi = {10.1088/1751-8121/acc7db},
  urldate = {2025-10-10},
  langid = {english}
}

@article{romanoDecoherenceBasedApproachClassical2023,
  title = {A {{Decoherence-Based Approach}} to the {{Classical Limit}} in {{Bohm}}'s {{Theory}}},
  author = {Romano, Davide},
  year = 2023,
  month = mar,
  journal = {Foundations of Physics},
  volume = {53},
  number = {2},
  pages = {41},
  issn = {1572-9516},
  doi = {10.1007/s10701-023-00679-w},
  urldate = {2025-10-29},
  langid = {english}
}

@article{schiffCommunicationQuantumMechanics2012,
  title = {Communication: {{Quantum}} Mechanics without Wavefunctions},
  shorttitle = {Communication},
  author = {Schiff, Jeremy and Poirier, Bill},
  year = 2012,
  month = jan,
  journal = {The Journal of Chemical Physics},
  volume = {136},
  number = {3},
  pages = {031102},
  issn = {0021-9606},
  doi = {10.1063/1.3680558},
  urldate = {2025-10-23}
}

@article{sebensFundamentalityFields2022,
  title = {The Fundamentality of Fields},
  author = {Sebens, Charles T.},
  year = 2022,
  month = sep,
  journal = {Synthese},
  volume = {200},
  number = {5},
  pages = {380},
  issn = {1573-0964},
  doi = {10.1007/s11229-022-03844-2},
  urldate = {2025-11-19},
  langid = {english}
}

@article{sebensQuantumMechanicsClassical2015,
  title = {Quantum {{Mechanics}} as {{Classical Physics}}},
  author = {Sebens, Charles T.},
  year = 2015,
  month = apr,
  journal = {Philosophy of Science},
  volume = {82},
  number = {2},
  pages = {266--291},
  issn = {0031-8248, 1539-767X},
  doi = {10.1086/680190},
  urldate = {2025-07-31},
  langid = {english}
}

@article{struyvePilotwaveApproachesQuantum2011,
  title = {Pilot-Wave Approaches to Quantum Field Theory},
  author = {Struyve, Ward},
  year = 2011,
  month = jul,
  journal = {Journal of Physics: Conference Series},
  volume = {306},
  number = {1},
  pages = {012047},
  issn = {1742-6596},
  doi = {10.1088/1742-6596/306/1/012047},
  urldate = {2026-01-30},
  langid = {english}
}

@article{takabayasiHydrodynamicalRepresentationNonRelativistic1954,
  title = {On the {{Hydrodynamical Representation}} of {{Non-Relativistic Spinor Equation}}},
  author = {Takabayasi, Takehiko},
  year = 1954,
  month = dec,
  journal = {Progress of Theoretical Physics},
  volume = {12},
  number = {6},
  pages = {810--812},
  issn = {0033-068X},
  doi = {10.1143/PTP.12.810},
  urldate = {2026-01-18}
}

@article{takabayasiRelativisticHydrodynamicsDirac1957,
  title = {Relativistic {{Hydrodynamics}} of the {{Dirac Matter}}. {{Part I}}. {{General Theory}}*},
  author = {Takabayasi, Takehiko},
  year = 1957,
  month = jan,
  journal = {Progress of Theoretical Physics Supplement},
  volume = {4},
  pages = {1--80},
  issn = {0375-9687},
  doi = {10.1143/PTPS.4.2},
  urldate = {2026-01-18}
}

@article{valentiniSignallocalityHiddenvariablesTheories2002,
  title = {Signal-Locality in Hidden-Variables Theories},
  author = {Valentini, Antony},
  year = 2002,
  month = may,
  journal = {Physics Letters A},
  volume = {297},
  number = {5},
  pages = {273--278},
  issn = {0375-9601},
  doi = {10.1016/S0375-9601(02)00438-3},
  urldate = {2025-10-15}
}

@book{wallaceEmergentMultiverseQuantum2012,
  title = {The {{Emergent Multiverse}}: {{Quantum Theory}} According to the {{Everett Interpretation}}},
  shorttitle = {The {{Emergent Multiverse}}},
  author = {Wallace, David},
  year = 2012,
  month = may,
  publisher = {Oxford University Press},
  address = {Oxford},
  doi = {10.1093/acprof:oso/9780199546961.001.0001},
  urldate = {2025-10-08},
  isbn = {978-0-19-954696-1}
}

@book{wyattQuantumDynamicsTrajectories2005,
  title = {Quantum {{Dynamics}} with {{Trajectories}}},
  author = {Wyatt, Robert E},
  year = 2005,
  series = {Interdisciplinary {{Applied Mathematics}}},
  volume = {28},
  publisher = {Springer-Verlag},
  address = {New York},
  doi = {10.1007/0-387-28145-2},
  urldate = {2026-01-18},
  copyright = {http://www.springer.com/tdm},
  isbn = {978-0-387-22964-5},
  langid = {english}
}

@article{Bohm2,
  title = {A Suggested Interpretation of the Quantum Theory in Terms of "Hidden" Variables. I},
  author = {Bohm, David},
  journal = {Phys. Rev.},
  volume = {85},
  issue = {2},
  pages = {166--179},
  numpages = {0},
  year = {1952},
  month = {Jan},
  publisher = {American Physical Society},
  doi = {10.1103/PhysRev.85.166},
  url = {https://link.aps.org/doi/10.1103/PhysRev.85.166}
}

@article{PARLANT20123,
title = "Classical-like trajectory simulations for accurate computation of quantum reactive scattering probabilities",
journal = "Computational and Theoretical Chemistry",
volume = "990",
pages = "3 - 17",
year = "2012",
note = "Chemical reactivity, from accurate theories to simple models, in honor of Professor Jean-Claude Rayez",
issn = "2210-271X",
doi = "https://doi.org/10.1016/j.comptc.2012.01.034",
url = "http://www.sciencedirect.com/science/article/pii/S2210271X1200059X",
author = "Gérard Parlant and Yong-Cheng Ou and Kisam Park and Bill Poirier"
}

@article{LiCaH,
author = {Scribano,Yohann  and Parlant,Gérard  and Poirier,Bill },
title = {Communication: Adiabatic quantum trajectory capture for cold and ultra-cold chemical reactions},
journal = {The Journal of Chemical Physics},
volume = {149},
number = {2},
pages = {021101},
year = {2018},
doi = {10.1063/1.5041091},

URL = { 
        https://doi.org/10.1063/1.5041091
    
},
eprint = { 
        https://doi.org/10.1063/1.5041091
    
}

}

@unpublished{rqtpap,
	Author = {Bill Poirier},
	Note = {\textit{ArXiv Preprint} 1208.6260 [quant-ph]},
	Title = {Trajectory-based Theory of Relativistic Quantum Particles},
	Year = {2012}}

@article{poirier24spin2,
	Author = {Bill Poirier and Richard Lombardini},
	Journal = {Entropy},
	Pages = {336},
	Title = {Dwell Times, Wavepacket Dynamics, and Quantum Trajectories for Particles with Spin 1/2},
	Volume = {26},
	Year = {2024}}

@article{poirier23nuclearoptpot,
	Author = {N. A. {Coleta da Concei\c{c}\~{a}o} and Brett V. Carlson and Bill Poirier},
	Journal = {Physica Scripta A},
	Pages = {115303},
	Title = {Quantum Trajectories and the Nuclear Optical Model},
	Volume = {98},
	Year = {2022}}

@article{poirier21fermi,
	Author = {Mahir S. Hussein and Bill Poirier},
	Journal = {Braz. J. Phys.},
	Pages = {{193--203}},
	Title = {Quantum Trajectory Description of the Time-Independent (Inverse) Fermi Accelerator,},
	Volume = {51},
	Year = {2021}}

@article{poirier20relsym,
	Author = {Bill Poirier and {Hung-Ming} Tsai},
	Journal = {Journal of Physics: Conference Series},
	Pages = {012022},
	Title = {Trajectory-based Conservation Laws for Massive Spin-zero Relativistic Quantum Particles in 1 + 1 Spacetime},
	Volume = {1612},
	Editor = {D. Schuch and M. Ramek},
	Publisher = {IOP Publishing},
	Year = {2020}}

@article{poirier14prx,
	Author = {Bill Poirier},
	Journal = {Phys. Rev. X},
	Pages = {040002},
	Title = {The Many Interacting Worlds Approach to Quantum Mechanics},
	Volume = {4},
	Year = {2014}}

@inbook{poirier11nowaveCCP6,
	Address = {Daresbury Laboratory},
	Author = {B. Poirier},
	Chapter = {Trajectory-Based Derivation of Classical and Quantum Mechanics},
	Editor = {K. H. Hughes and G. Parlant},
	Pages = {6},
	Publisher = {CCP6},
	Title = {Quantum Trajectories},
	Year = {2011}}

@article{bostrom2015,
  author  = {Bostr{\"o}m, Kim Joris},
  title   = {Quantum mechanics as a deterministic theory of a continuum of worlds},
  journal = {Quantum Studies: Mathematics and Foundations},
  year    = {2015},
  volume  = {2},
  number  = {4},
  pages   = {315--347},
  doi     = {10.1007/s40509-015-0046-6}
}

\end{document}